\documentclass[11pt]{article}
\usepackage{graphicx}
\usepackage[latin2]{inputenc}
\usepackage{graphicx}
\usepackage{amsfonts}
\usepackage{amssymb}
\usepackage{amsmath}
\usepackage{newclude}
\usepackage{appendix}
\usepackage{authblk}
\usepackage[square,comma,compress,numbers]{natbib}
\usepackage{hyperref}
\usepackage{float}
\usepackage{tikz}
\usepackage{gastex}

\usetikzlibrary{through}
\usetikzlibrary{intersections}
\usetikzlibrary{calc}
\usepackage{xcolor}
\usepackage{adjustbox}
\usepackage{psfrag}
\usepackage{authblk}
\usepackage{breqn}

\newcommand{\tikzAngleOfLine}{\tikz@AngleOfLine}
  \def\tikz@AngleOfLine(#1)(#2)#3{%
  \pgfmathanglebetweenpoints{%
    \pgfpointanchor{#1}{center}}{%
    \pgfpointanchor{#2}{center}}
  \pgfmathsetmacro{#3}{\pgfmathresult}%
}

\newtheorem{Def}{Definition}

\DeclareMathOperator*{\acosh}{acosh}

\newcommand{\Exp}{{\rm Exp}}

\makeatletter
\renewcommand{\@dotsep}{10000}
\makeatother

\begin{document}

\graphicspath{{./paper/}}
\title {Navigable Networks as Nash Equilibria of Navigation Games}
\author{%
  Andr\'as Guly\'as$^{1,3}$, %
  J\'ozsef J. B\'ir\'o$^{1,2}$, %
  Attila K\H{o}r\"osi$^{3}$, %
  G\'abor R\'etv\'ari$^{1,2}$ \& %
  Dmitri Krioukov$^{4}$ \\
  $^{1}$ High Speed Networks Laboratory, Department of
  Telecommunications and Media Informatics, Budapest University of Technology
  and Economics, 2 Magyar tud\'osok str., Budapest 1117, Hungary. \\
  $^{2}$ MTA-BME Future Internet Research Group, Budapest University of
  Technology and Economics, 2 Magyar tud\'osok str., Budapest 1117,
  Hungary.\\
  $^{3}$ MTA-BME Information Systems Research Group, Budapest University
  of Technology and Economics, 2 Magyar tud\'osok str., Budapest 1117,
  Hungary. \\
  $^{4}$ Northeastern University, Department of Physics, Department of
  Mathematics, Department of Electrical\&Computer Engineering, 360 Huntington
  Ave, 111 Dana Research Center, Boston, MA 02115, USA. 
}

\date{}
\maketitle

\begin{abstract}

  The common sense suggests that networks are not random mazes of
  purposeless connections, but that these connections are organised so
  that networks can perform their functions well. One function common
  to many networks is targeted transport or navigation. Using game
  theory, here we show that minimalistic networks designed to maximise
  the navigation efficiency at minimal cost share basic structural
  properties with real networks. These idealistic networks are Nash
  equilibria of a network construction game whose purpose is to find
  an optimal trade-off between the network cost and navigability.  We
  show that these skeletons are present in the Internet, metabolic,
  English word, US airport, Hungarian road networks, and in a
  structural network of the human brain. The knowledge of these
  skeletons allows one to identify the minimal number of edges by
  altering which one can efficiently improve or paralyse navigation in
  the network.

\end{abstract}

Networks are efficient conduits of information and other media. News,
ideas, opinions, rumours, and diseases spread through social networks
fast, sometimes becoming viral for reasons that are often difficult to
predict
\cite{barrat2008dynamical,kitsak2010identification,watts2002identity,pastor2001epidemic,rhodes1996power,ferguson2007capturing,doerr2013lognormal,meloni2009traffic,barthelemy2004velocity,miritello2011dynamical,gallos2007scaling,moreno2004dynamics}. Many
biological networks are also paradigmatic examples of information
routing, ranging from information processing and transmission in the
brain, to signalling in gene regulatory networks, metabolic networks,
or protein
interactions~\cite{BaOl04,YaBo09,BuSp09,Chialvo2010}. Perhaps the most
basic example is the Internet whose primary function is to route
information between computers. If one is to list some common functions
of different networks, then information routing will likely be close
to the top. It is thus not surprising that many networks were found
navigable, meaning that nodes can efficiently route information
through the network even though its global structure is not known to
any individual
node~\cite{milgram67,TraMi69,kleinberg2000navigation,DoMuWa03,LiNoKu05,SiJe08,BoKrKc08,CaRi09,HuWa11,LeHo12,LeHo12a,NSW_Yang_2015,CaBo12}.

These findings do not necessarily mean that real networks evolve to
become navigable. Navigability can be a by-product of some other
evolutionary incentives because different networks have many other
different functions as well. In other words, it remains unclear if
ideal networks whose only purpose is to be maximally navigable at
minimal costs have anything in common with real networks. Even if they
do, then how close are real networks to these ideal maximally
navigable configurations? If they are close but not exactly there, or
if their navigability suddenly deteriorates, possibly signifying an
onset of a disease~\cite{CoLeMo11}, then what can we do to cure the
network and boost its navigability?

Here we show that the ideal maximally navigable networks do share some
basic structural properties with the Internet, {\it E.coli} metabolic
network, English word network, US airport network, the Hungarian road
network, and a structural network of the human brain.  Yet these ideal
networks are not generative models of the real networks, where by
generative models we mean function-agnostic models that simply try to
reproduce some structural properties of real networks.  Instead these
ideal networks identify minimal sets of edges that are most critical
for navigation in the real network.  In other words, they are
navigation skeletons or subgraphs of real networks.  We find that the
considered real networks contain high percentages, exceeding $90\%$ in
certain cases, of edges from their navigation skeletons, while the
probability of such containment in randomized null models is
exponentially small.  The knowledge of these skeletons allows us to
quantify exactly what connections the considered real networks lack to
be maximally navigable, and which of their connections are not exactly
necessary for that. To define and construct these maximally navigable
network skeletons we employ game theory.

\begin{figure}[ht]
  \centering
  \includegraphics[width=0.45\textwidth,angle=0]{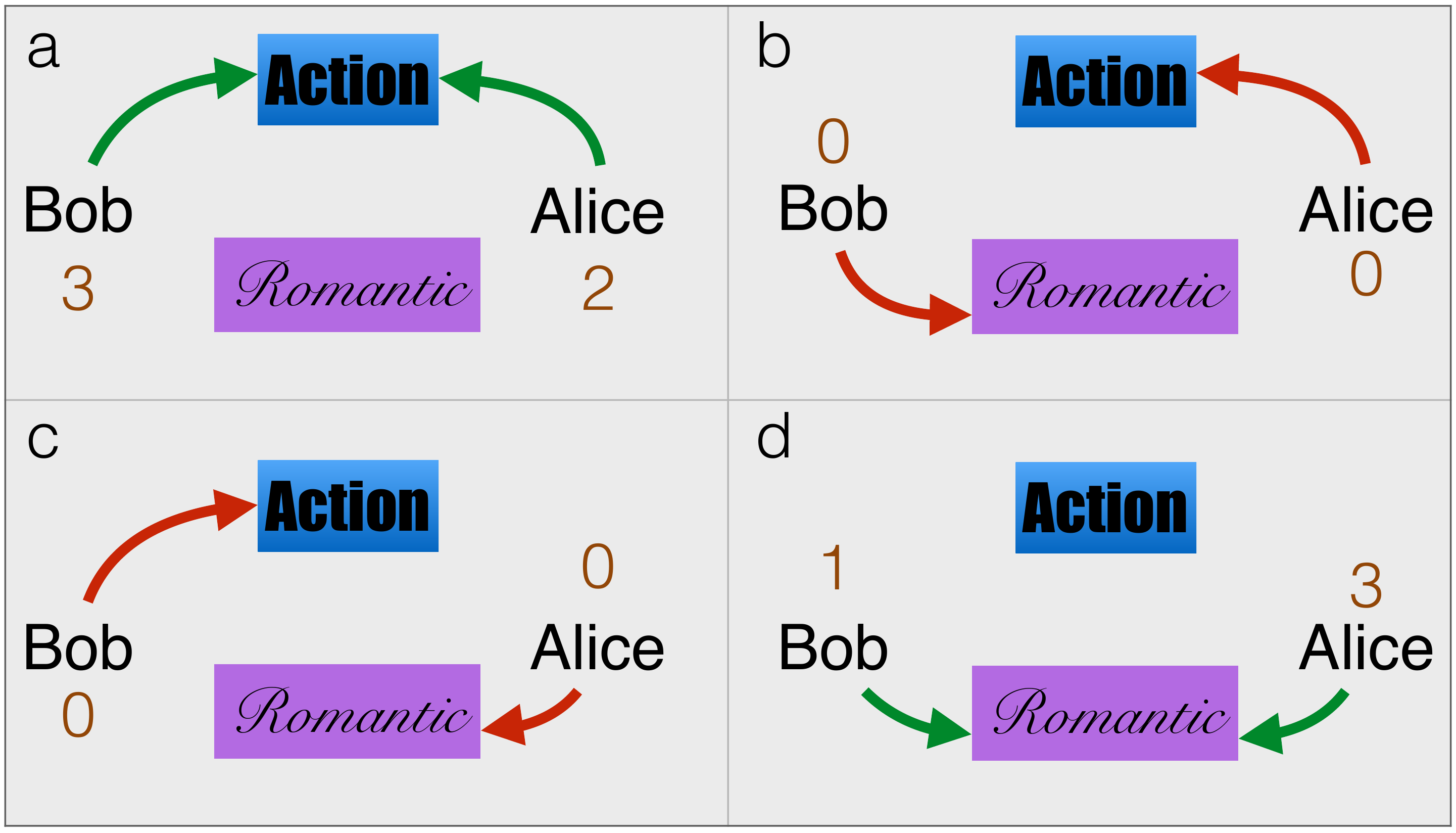}
  \caption{Illustration of game theory. Alice and Bob are
    happy only if they go out to the movies together, but the level of
    their happiness depends on what movie they watch. The basic
    notions of a game: Players: Alice and Bob; Strategies: Go to see
    an action or a romantic movie; Payoffs: The level of happiness
    $0,1,2,3$; Nash equilibria: situations in which the players cannot
    be happier by unilaterally modifying their strategies. In the
    figure, states~(a) and~(d) are equilibria when Alice and Bob go
    together to watch a movie. State~(a) is the global optimum since
    the total happiness $3+2=5$ is maximised. \label{fig:1}}
\end{figure}

Game theory is a standard tool to study the behaviour of a population
with given incentives. The population members are called players, and
their possible actions are strategies, while cost functions or payoffs
express players' incentives. The purpose of a player is to minimise
her costs (or maximise her payoffs) by adjusting her strategy. A Nash
equilibrium is a game state such that no player can further reduce her
costs by altering her strategy unilaterally. Such equilibrium states
are local optima where the game can eventually settle after some
transient dynamics.  The global optimum is an optimum where the total
cost of all players is minimised. Since the inception of game theory a
broad palette of games has been introduced, modelling diverse
properties of real-life situations~\cite{nisan2007algorithmic},
Figure~\ref{fig:1}.

Here we use game theory to find the structure of networks that are
Nash equilibria of a network construction game
\cite{nisan2007algorithmic,Fabrikant:2003:NCG:872035.872088,
  Anshelevich04theprice, Corbo:2005:PSB:1073814.1073833,
  Albers06onnash, Demaine:2007:PAN:1281100.1281142, mihalak2010price}
with navigability incentives.  The concept of Nash equilibrium
captures the idea of self-organisation, i.e., of the emergence of
structures from the local interaction of rational but selfish players,
in contrast with global optimisation used in centralised planning of
globally optimal navigable structures~\cite{lee2013greedy}. In our
Network Navigation Game~(NNG), players are network nodes whose optimal
strategy is to set up a minimal number of edges to other nodes
ensuring maximum navigability.  That is, the cost function reflects
trade-offs between the number of created edges and navigability.  If
each node connects to each other node, then this construction is
maximally navigable but maximally expensive, too. If no edges are set
up, then the cost is zero, but so is navigability. There is a sweet
spot of the least expensive but still $100\%$-navigable network,
defined as the network in which all pairs of nodes can successfully
communicate using geometric routing~\cite{PaRa05}.  The goal of our
game is to find this sweet spot.

\section{Results}

The network construction game that we employ is very general and applies to any set
of points in any geometry. The latent geometry of numerous real
networks is not Euclidean but hyperbolic as shown in~\cite{PaBoKr11}.
Specifically, the model in~\cite{PaBoKr11} extends the preferential
attachment mechanism of network growth by observing that in many real
networks the probability of establishing a connection depends not only
on popularity of nodes, i.e., their degrees, but also on similarity
between nodes. Similarity is modeled in~\cite{PaBoKr11} as a distance
between nodes on the simplest compact space, the circle. The
connection probability thus depends both on node degrees (popularity)
and on the distance between nodes on the circle (similarity). The node
degrees are then mapped to radial coordinates of nodes, thus moving
nodes from the circle to its interior, the disk. One can then show
that the resulting connection probability depends only on the
hyperbolic (versus Euclidean) distance between nodes on the disk, and
that the resulting graphs are random geometric
graphs~\cite{Penrose03-book} growing over the hyperbolic plane. As
shown in earlier work~\cite{KrPa10}, these graphs are maximally
random, i.e., maximum-entropy graphs that have power-law degree
distributions and strong clustering. In other words, power-law degree
distributions, coupled with strong clustering, are manifestations of
latent hyperbolic geometry in networks. If this geometry is not
hyperbolic but Euclidean, then the resulting random geometric graphs
still have strong clustering, but their degree distributions are
Poisson distributions that do not have any fat
tails~\cite{Penrose03-book}.  The model in~\cite{PaBoKr11} has been
validated against long histories of growth of several real networks,
predicting their growth dynamics with a remarkable precision. It is
then not surprising that as a consequence the same model also
reproduces a long list of structural properties of these
networks~\cite{PaBoKr11}.

Random geometric graphs~\cite{Penrose03-book} are defined as sets of
points sprinkled uniformly at random over a (chunk of) geometric
space. Every pair of points is then connected if the distance between
the points in the space is below a certain threshold.  Given that the
latent space of real scale-free networks is hyperbolic, our starting
point is the first part (uniform sprinkling) of the random geometric
graph definition. That is, we first randomly sprinkle a set of points
over a hyperbolic disk. We then do not proceed to the second part of the
random geometric graph definition. Instead, given only the coordinates
of sprinkled nodes, we identify the sets of edges, ideal for
navigation, that correspond to the Nash equilibria of our NNGs. We
then analyse the structural properties of the resulting
ideal-navigation networks, and find that, surprisingly, they also have
power-law degree distributions and strong clustering. This result
invites us to investigate if these navigation-critical edges exist in
real networks. To check that, we have to know the hyperbolic
coordinates of nodes in these real networks in the first place. We
infer these coordinates in the considered collection of real networks
using the deterministic HyperMap algorithm (Methods).  Given only
these inferred coordinates, we then construct the ideal-navigation
Nash equilibria defined by these coordinates, and compare, edge by
edge, the resulting Nash equilibrium networks against the real
networks.  We find that the real networks contain large percentages of
edges from their Nash equilibria. This methodology thus allows us to
identify the navigation skeleton of a given real network.  We finally
check directly that edges in these skeletons are indeed most critical
for navigation by showing that their alterations affect drastically
network navigability.

\noindent \textbf{Game definition.}

%\noindent Players, hyperbolic plane, and geometric routing. \hspace{.5cm}
We start
with a set of players $u=1,2,\ldots,N$, i.e., $N$ nodes, scattered
randomly over a hyperbolic disk of radius~$R$. The densities of
players' polar coordinates~$(r,\phi)$, $r\in[0,R]$, $\phi\in[0,2\pi]$,
are~\cite{KrPa10}
\begin{equation}\label{eq:densities}
  \rho(r) = \frac{\alpha\sinh(\alpha r)}{\cosh(\alpha R)-1},\quad\rho(\phi)=\frac{1}{2\pi},
\end{equation}
where $\alpha>1/2$ is a parameter controlling the heterogeneity of the
layout. If $\alpha=1$, the players are distributed uniformly over the
hyperbolic disk because the area element at coordinates~$(r,\phi)$ is
$dA=\sinh(r)\,dr\,d\phi$. The desired player scattering is achieved in
simulations by placing players~$u$ at polar coordinates
$r_u=(1/\alpha)\acosh\left\{1+\left[\cosh(\alpha R)-1\right]U\right\}$
and $\phi_u=2\pi U$ where $U$ for each~$u$ is a random number drawn
from the uniform distribution on $[0,1]$. The hyperbolic distance
between any two players $u$ and $v$ is
\begin{equation}
  d(u,v)=\acosh\left[\cosh r_u\cosh r_v-\sinh r_u\sinh r_v\cos(\phi_u - \phi_v)\right].
\end{equation}
In greedy geometric routing, player~$u$ routes information to some
remote player~$v$ by forwarding the information to its connected
neighbour~$u'$ closest to~$v$ in the plane according to the distance
above.  If $u$ has no neighbour~$u'$ closer to~$v$ than $u$self, then
navigation fails, and we say that~$u$ cannot navigate to~$v$.  The
percentage of pairs of players~$u,v$ such that $u$ can successfully
navigate to~$v$ is called the success ratio.  If this percentage
is~$100\%$, we say that the network is maximally ($100\%$) navigable.

%\noindent Strategies. \hspace{.5cm}
The strategy space of player~$u$ is all possible
combinations of edges that $u$ can establish to other players.  One
extremal strategy is to establish no edges. The other extreme is to
connect to everyone.  The total number of possibilities for~$u$ is
$2^{N-1}$. Any combination of strategies that all players select is a
network on $N$ nodes.

%\noindent Payoff.  \hspace{.5cm}
The objective of each player~$u$ is to set up a
minimal number of edges to other players such that~$u$ can still
navigate to any other player in the network. Formally, the cost
function of player $u$ that it minimises is $c_u=k_u+n_u$, where $k_u$
is the number of edges that $u$ establishes, and $n_u$ is either zero
if $u$ can navigate to everyone, or infinity otherwise. A more formal
description of the strategies and payoffs can be found in
Appendix~1.

\noindent \textbf{Nash equilibria of the game.}

\begin{figure}[H]
  \centering
  \includegraphics[width=0.99\textwidth,angle=0]{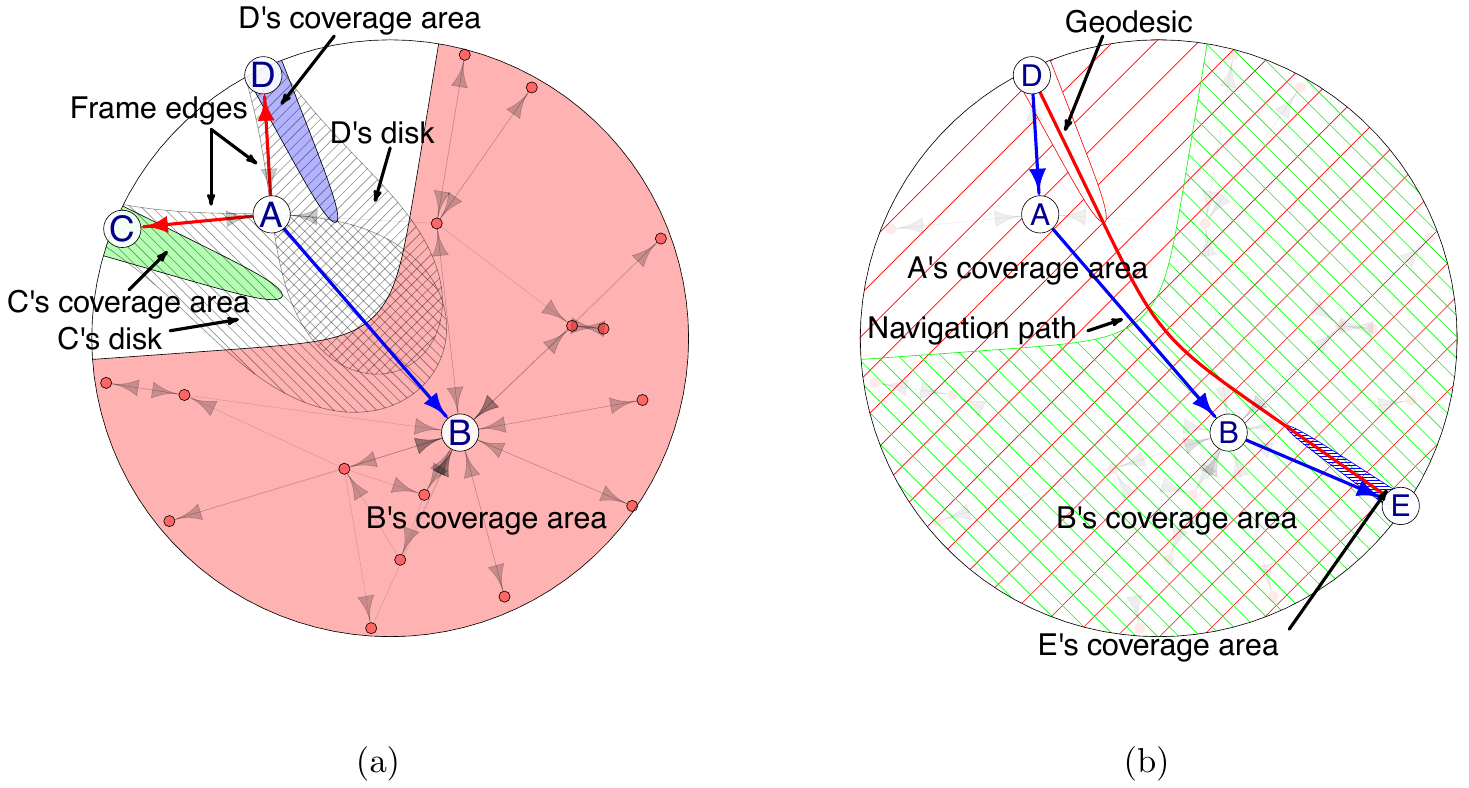}
  \caption{Illustration of the network navigation game (NNG). Panel
    (a) shows the optimal set of connections (optimal strategy) of
    node A in a small simulated network.  All nodes are distributed
    uniformly at random over the hyperbolic disk, and A's optimal
    strategy is to connect to the smallest number of nodes ensuring
    maximum (100\%) navigability.  These nodes are B, C, and D because
    it is the smallest set of nodes whose coverage areas, shown by the
    coloured shapes, contain all other nodes in the network. B's
    coverage area for~A (red) is defined as a set of points
    hyperbolically closer to B than to A, therefore if A is to
    navigate to any point in this area, A can select B as the next
    hop, and the message will eventually reach its destination, as the
    second panel illustrates.  Link AC (and AD) in panel (a) is also a
    frame link, because A is the closest node to~C, as illustrated by
    the hyperbolic disk of radius~$|AC|$ centred at~C (the line-filled
    shape), which does not contain any nodes other than C and
    A. Therefore to navigate to~C, A has no choice other than to
    connect directly to~C.  Panel (b) shows the sequence of shrinking
    coverage areas along the navigation path (blue arrows) from~D
    to~E. The red curve is the geodesic between~D and~E in the
    hyperbolic plane.  The coverage areas are shown by the shapes
    filled with lines of increasing density. The largest is A's
    coverage for~D. The next one is B's coverage for~A. The smallest
    is E's coverage for~B. \label{fig:2}}
\end{figure}

Given any player~$u$, we call player $v$'s coverage area the set of
all points closer to~$v$ than to~$u$, Figure~\ref{fig:2}.  Trivially $v$
covers itself, since it is closer to itself ($d(v,v)=0$) than
to~$u$. Therefore if $u$ connects to all other players, then $u$
trivially covers them all. The optimal strategy for~$u$ minimising
$u$'s costs is thus to connect to a minimal number of players such
that the union of their coverage areas contains all the other
players. Indeed, if $u$ does that, and if all other players do the
same, then the resulting network is $100\%$-navigable at minimal
number of edges. The network is fully navigable because if $u$ wants
to navigate to any remote player~$w$, then by construction there
exists $u$'s neighbour~$v$ that contains~$w$ in its coverage area, and
$u$ can use~$v$ as the next hop towards~$w$. If $v$ is not directly
connected to~$w$, then there exists $v$'s neighbour~$v'$ that
contains~$w$ in its coverage area, so that $v$ can route to~$v'$, and
so on until the information reaches destination~$w$ lying within the
intersection of all the coverage areas along the path,
Figure~\ref{fig:2}. The problem of finding the optimal set of edges
for~$u$ thus reduces to the minimum set cover problem
\cite{garfinkel1972integer}. A formal description of the equilibrium
network and a detailed example (for simplicity in the Euclidean plane)
can be found in Appendix~2.

The Nash equilibrium of this game is not necessarily unique. There can
exist different networks minimising the cost defined above.  As
specified in Appendix~2, in what follows, among all the NNG
equilibria, we always select the unique one that minimises the sum of
distances span by its edges, thus making the NNG Nash equilibrium
network construction deterministic.  However there also exist certain
edges, which we call frame edges, necessarily present in any Nash
equilibrium. Edge~$u \to v$ is a frame edge if $u$ is the closest
player to~$v$. In this case $u$ cannot navigate to $v$ through any
other players since there is no one closer to~$v$ than $u$self, so
that $u$ must connect directly to~$v$ to reach it,
Figure~\ref{fig:2}. If at least one of such edges is absent, the network
is not fully navigable.  The exact definition of the ``frame
topology'' consisting the frame edges can be found in Appendix~3.

In any Nash equilibrium of this game, each player computes its optimal
strategy independently of others. In game theory such equilibria are
called dominant strategy equilibria.  Moreover the equilibrium is also
a social optimum since one cannot create a fully navigable network
using less edges.

\noindent \textbf{Structural properties of Nash-equilibrium networks.}

Using the trigonometry of overlapping hyperbolic disks, we show in
Appendix~4 that if the node density is uniform ($\alpha=1$), then the
probability $p(d)$ that two players $u$ and $v$ located at distance
$d\equiv d(u,v)$ are connected in a Nash equilibrium network lies
between $\exp(-8\, \delta \, e^{d/2})$ and $\exp(-2 \, \delta \,
e^{d/2})$,
\begin{equation}
e^{-8\, \delta \, e^{d/2}} \leq p(d) \leq e^{-2\, \delta \, e^{d/2}} \ ,
\end{equation}
where $\delta$ is the average density of players on the disk, that is $\delta = N/A$, where $A$ is the disk area.
The expected degree of player~$u$ at polar coordinates~$(r_u,0)$---we can assume that $u$'s angular coordinate is $\phi_u=0$ without loss of generality---is then
$\bar{k}(r_u)= N \int p[d(u,v)]\,\rho(r_v)\rho(\phi_v)\,dr_v\,d\phi_v$, where $\rho(r_v)$ and $\rho(\phi_v)$
are the player densities from Eq.~(\ref{eq:densities}). We can
evaluate this integral to find that the expected number~$\bar{k}(r)$
of connections of a player at radial coordinate~$r$ is bounded by
(analytically shown in Appendix~5)
\begin{equation}\label{eq:k(r)-bounds}
\frac{1}{2}e^{(R-r)/2}\leq \bar{k}(r)\leq2\,e^{(R-r)/2} \ ,
\end{equation}
where $r\equiv r_u$. It then follows that the average degree of players in the network, given by $\bar{k}=\int_0^R\bar{k}(r)\rho(r)\,dr$, lies between $1$~and~$4$,
\begin{equation}\label{eq:ave-deg}
1\leq\bar{k}\leq4.
\end{equation}
We also see from Eq.~(\ref{eq:k(r)-bounds}) that the degree of players
decays exponentially as the function of their radial position,
$\bar{k}(r)\sim e^{-r/2}$, while their density exponentially
increases, $\rho(r)\sim e^r$, Eq.~(\ref{eq:densities}). The
combination of these two exponentials yields the power-law degree
distribution (see Appendix 6 for the detailed derivation) in the
network~\cite{BoPa03,newman05} \begin{equation}
  P(k)=\frac{1}{k!}\int_0^Re^{-\bar{k}(r)}[\bar{k}(r)]^k\rho(r)\,dr=2\left(\frac{\bar{k}}{2}\right)^2\frac{\Gamma(k-2,\bar{k}/2)}{k!}\sim
  k^{-3}.
\end{equation}
We also show analytically in Appendix~7-8, that the average
clustering $\bar{c}(k)$ of players of degree~$k$ decays with~$k$ as
$1/k$, while the average clustering $\bar{c} = \sum_k P(k)\bar{c}(k)$
in the network is around $0.45$, also confirmed in
simulations. Clustering does not depend on network size or average
degree, meaning that clustering is a positive constant even in the
large graph size limit. Remarkably, neither degree distribution nor
clustering depend on the player density~$\delta$.

For non-uniform node density $\alpha\neq1$, we can analytically obtain
only the lower bound for $\bar{k}(r,\alpha)$, which is still
proportional $e^{-\frac{r}{2}}$, i.e., independent of~$\alpha$ if
$\alpha>1/2$, Appendix~9. This
lower bound suggests that the degree distribution is a power law $P(k)
\sim k^{-\gamma}$ with exponent $\gamma=2\alpha+1$, which we confirm
in simulations in Appendix~9.  Figure~\ref{fig:4} shows that
the closer the $\gamma$ to~$2$, the stronger the clustering, the
cheaper the network, and the more efficient and robust the
navigability. The value of $\gamma=2$ thus appears as the ``best
choice'' for a network---the network is maximally navigable at the
lowest cost. These results complement existing
works~\cite{cohen2003scale,BoKrKc08} showing that $\gamma=2$ yields
most navigable networks, by adding that this $\gamma$ also provides a
minimum cost equilibrium topology as well, explaining the emergence of
these networks from the interaction of selfish players.

\begin{figure}[H]
  \centering
  \includegraphics[width=0.45\textwidth,angle=0]{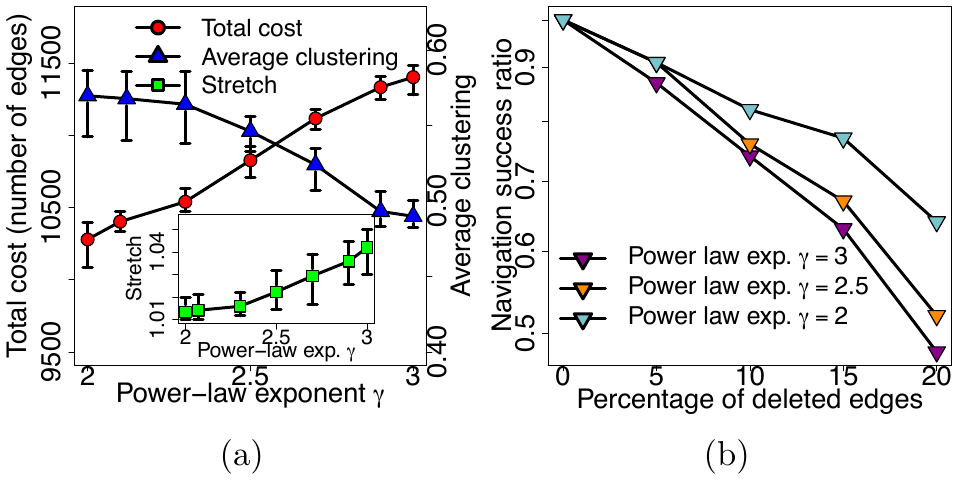}
  \caption{Topological properties of NNG equilibrium networks as a
    function of the power-law exponent. Panel (a) shows the total cost
    (number of edges), average clustering~$\bar{c}$, and stretch in
    NNG-simulated networks as functions of~$\gamma$. Stretch (shown in
    the inset) is the average factor showing by how much longer the
    greedy navigation paths are, compared to the shortest paths in the
    network. Stretch equal to~$1$ means that all navigation paths are
    shortest possible. The plotted points are mean values while the
    error bars show minimum and maximum values obtained for the NNG
    over 10 random sprinkling of nodes for a given value of $\gamma$.
    Panel (b) shows the success ratio as a function of the percentage
    of edges randomly deleted from the network. The smaller
    the~$\gamma$, the more robust the navigability with respect to
    this network damage. \label{fig:4}}
\end{figure}

\begin{figure}[H]
  \centering
  \includegraphics[width=0.45\textwidth,angle=0]{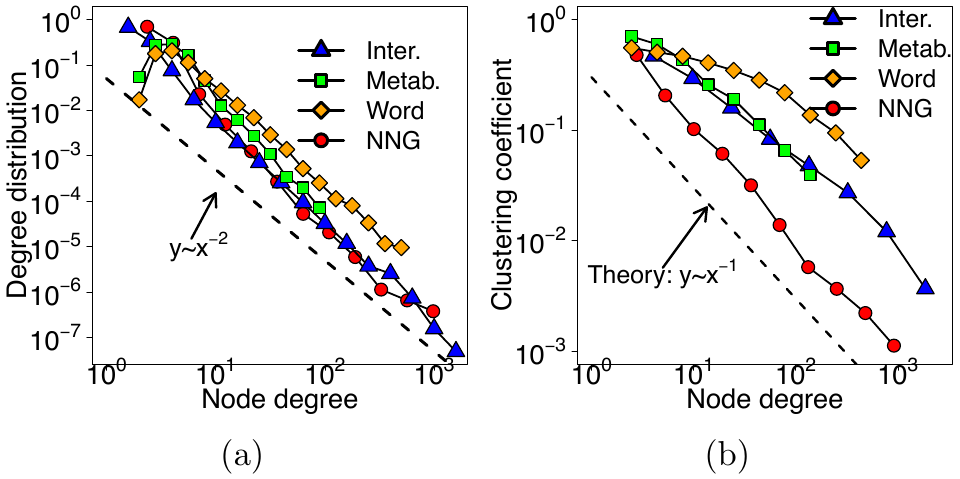}
  \caption{NNG equilibrium networks share basic structural properties
    with real networks. The real networks considered are the Internet,
    metabolic reactions, and the English word network, see
    Methods. Panel (a) and (b) shows the degree distribution and the
    average clustering coefficient of nodes of a given degree in the
    real and NNG networks. The dashed black lines are the power laws
    with exponents $-2$ and $-1$.  The power law decay of the
    clustering coefficient for the NNG is shown analytically in
    Appendix 7. The clustering coefficient of a node of
    degree $k$ is the number of triangular subgraphs containing the
    node, divided by the maximum possible such number, which is
    $k(k-1)/2$. In the NNG network, the disk radius is $R=21.2$ and
    $\alpha=0.5$. There are no other parameters.  \label{fig:3}}
\end{figure}

Figure~\ref{fig:3} and Table~\ref{tab:comp} confirm our analytic
results and shows that some basic structural properties of
NNG-simulated networks are similar to some real networks. 
\begin{table}[H]
  \centerline{
    \begin{tabular}{|c|c|c|c|c|}
      \hline
      Network & Inter.\ & Metab.\ &  Word & NNG\\
      \hline
      Nodes & 23748 & 602 & 4065 & 5000\\
      \hline
      Edges & 58414  & 2498 & 38631 & 7955\\
      \hline
      Avg.\ deg.\ & 4.92 & 8.29 & 19.01 & 3.18\\
      \hline
      Avg. clust.\ & 0.61 & 0.55 & 0.45 & 0.60\\
      \hline
      Avg.\ dist.\ & 3.52 & 3.22 & 2.43 & 3.89\\
      \hline
      Diam.\ & 10 & 6 & 6 & 10\\
      \hline
    \end{tabular}
  }
  \caption{ Comparison of basic structural properties of real and NNG
    networks. The average
    distance and diameter are the average and
    maximum hop lengths of the shortest paths in the network. The
    average degree in the NNG-simulated network is lower than in the
    real networks because the NNG generates navigable networks with
    minimum numbers of edges. In the NNG network, the disk radius is
    $R=21.2$ and $\alpha=0.5$. There are no other parameters. }
  \label{tab:comp}
\end{table}

Our results
also suggest that the incentive for navigability alone may be
sufficient to explain the properties of complex networks to a certain
degree. Yet we cannot really make this claim based only on such
large-scale statistical similarities. A more detailed link-by-link
comparison between real and corresponding NNG networks is needed to
understand how well the NNG reflects reality.

\noindent \textbf{Network Navigation Game versus real networks.}

\begin{figure}[H]
  \centering
  \includegraphics[width=0.75\textwidth,angle=0]{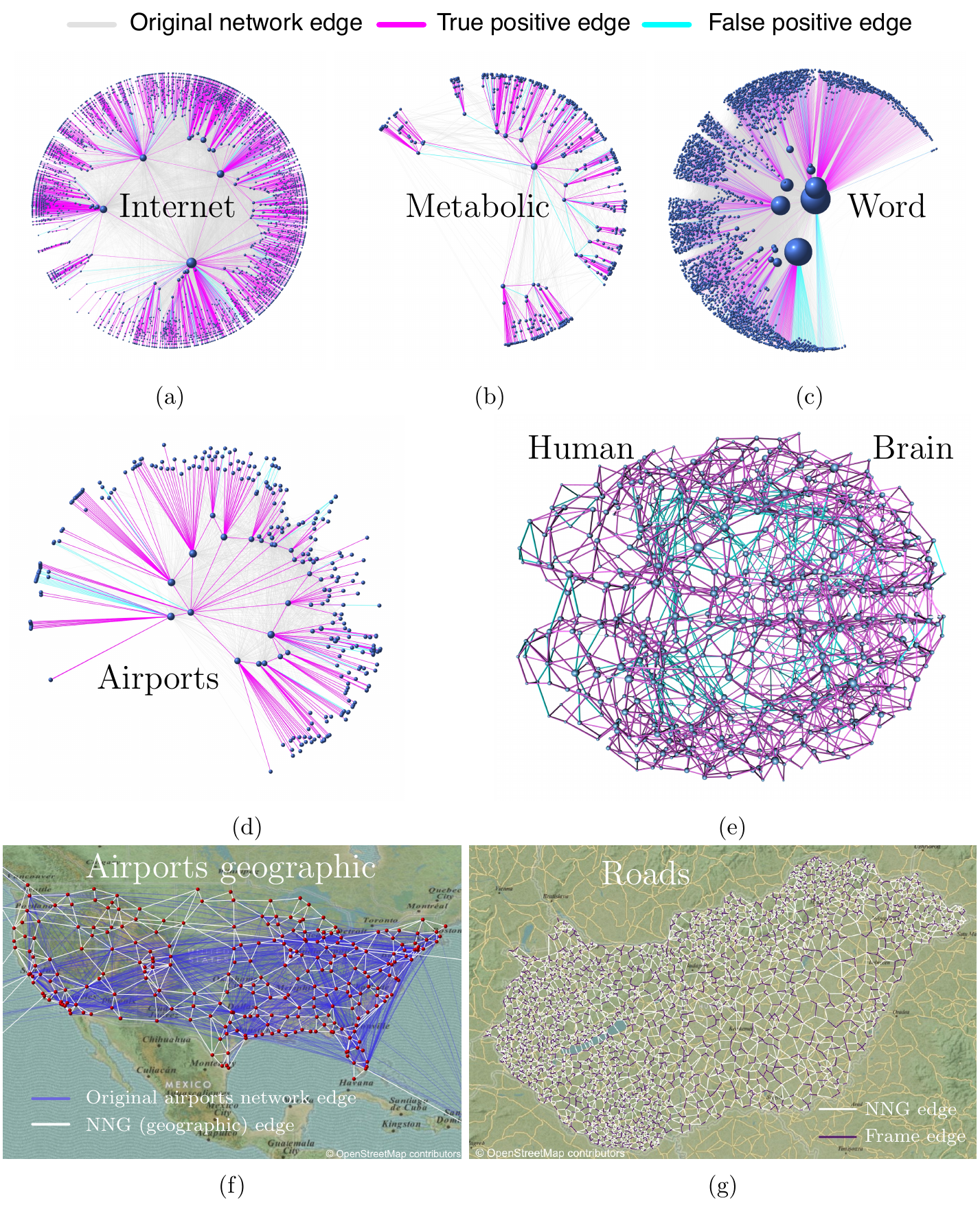}
  \caption{ \small Network Navigation Game~(NNG) predicts well links
    in real networks. Panels (a), (b), and (c) visualise the Internet,
    metabolic, and word networks mapped to the hyperbolic plane as
    described in the Methods section. The hyperbolic coordinates of
    nodes are then supplied to the minimum set cover algorithm that
    finds a Nash equilibrium of the NNG for each network.  Panels (d)
    and (e) do the same for the US airport network and for the human
    brain, except that in the brain the physical coordinates of nodes
    are used. The grey edges are present in the real networks but not
    in the NNG networks. These edges may exist in real networks for
    different purposes other than navigation, so that the NNG can say
    nothing about them. The false positive turquoise edges are present
    in the NNG networks but not in the real networks.  The true
    positive magenta edges are present in both networks. Panels (f)
    and (g) show the NNG equilibrium network based on the physical
    (geographic, versus hyperbolic) coordinates of US airports, and
    the NNG network for the Hungarian road network.  The NNG networks
    have the same sets of nodes as the corresponding real networks,
    but the sets of edges are different. For visualisation purposes
    the grey edges are suppressed in the human brain and Hungarian
    road networks. The detailed statistics of edges are in
    Table~\ref{tab:tab1}. The cartography in the background of panels
    (f) and (g) are licensed as CC BY-SA (see www.openstreetmap.org/copyright
    for details).  \label{fig:5}}
\end{figure}

Figure~\ref{fig:5} and Table~\ref{tab:tab1} show the results of this
analysis applied to these and other real networks. 
\begin{table}[H]
\centerline{
  \small
  \begin{tabular}{|c|c|c|c|c|c|c|c|}
     \hline
     & Inter. H & Metab. H & Word H & Roads E & Airp. S & Airp. H & Brain E \\
     \hline
     Nodes & $4919$ & $602$ & $4065$ & $3136$ & $283$ & $283$ & $998$\\
     \hline
     Real edges ($|R|$) & $28361$  &  $2498$ & $38631$ & - &  $1973$ &
     $1973$ & $17865$\\
     \hline
     NNG edges ($|M|$) & $5490$  & $743$ & $4634$ & $9808$  & $643$ &
     $328$ & $2591$\\
     \hline
     True positives ($|T|$) & $4556$  & $643$ & $3311$ & $8776$  &
     $65$ & $277$ & $2306$\\
     \hline
     False positives ($|F|$) &
     $934$  & $100$ &
     $1323$ & $1032$  &
     $578$ & $51$ & $285$\\
     \hline
\bf  Precision ($|T|/|M|$) & $83\%$  &  $87\%$ & $71.5\%$ & $89.48\%$
&  $10.1\%$ & $84\%$ & $89\%$\\
     \hline
     Frame edges ($|M_F|$) & $3680$  &  $415$ & $3304$ & $3105$  &
     $199$ &  $249$ & $716$\\
     \hline
     \!\!\!Frame true positives ($|T_F|$) \!\!\!\!\! & $3243$  &  $378$ & $2528$ &
     $2931$  &  $15$ &  $216$ & $677$\\
     \hline
\bf   \!\!\!Frame prec. ($|T_F|/|M_F|$) \!\!\!\!\! & $88\%$  &  $91\%$ & $77\%$ &
$94.40\%$  &  $7.5\%$ &  $87\%$ & $94.6\%$\\
     \hline
     Navigation success ratio & $87\%$  &  $85\%$ & $81\%$ & -  &
     $54\%$ &  $89\%$ & $89\%$\\
     \hline
   \end{tabular}
   }
   \caption{The table quantifies the relevant edge statistics in Figure~\ref{fig:5}, showing
     the total number of edges in the real networks~$|R|$, and in their
     NNG equilibrium networks~$|M|$, the number of true positive (magenta edges in Figure~\ref{fig:5}) $|T|=|M\cap
     R|$, the number of false positive (turquoise edges in Figure~\ref{fig:5}) $|F|=|M\setminus R|$, and the true positive rate, or
     precision, defined as~$|T|/|M|$. The precision statistics is also shown for the frame edges.
     Capital letters H,E,S after the network names refer to the embedding geometry: H:hyperbolic, E:Euclidean, S:spherical. The Euclidean coordinates in the brain are three-dimensional.
    \label{tab:tab1}
  }
\end{table}
We cannot expect
real networks to be identical to NNG networks because the latter are
minimum-cost maximum-navigation idealisations, while each individual
real network performs many other functions different from
navigation. In particular, since real networks must be error-tolerant
and robust with respect to different types of network damage, we
expect the number of edges in real networks to be noticeably larger
than in their minimalistic NNG counterparts---something we indeed
observe in Table~\ref{tab:tab1}.  Yet if navigation efficiency does
matter for real networks, then we should expect a majority of edges
present in these NNG idealisations to be also present in the
corresponding real networks. Table~\ref{tab:tab1} confirms these
expectations as well. The NNG precision in predicting links in real
networks, defined as the ratio of NNG true positive links to the total
number of NNG links, exceeds $80\%$ for most networks, while the
precision in predicting frame links, crucial for navigation, exceeds
$90\%$ for some networks. In Appendix~10 we juxtapose these
numbers against the corresponding numbers in randomized null models,
where they are exponentially small, upper bounded by $0.1\%$. We also
note that since the real networks have many more links than NNG
networks, their navigability may not suffer much from missing a small
percentage of NNG links, as confirmed by the success ratio results in
the same figure.

Of particular interest to us here are networks that
are explicitly embedded in the physical space. In these cases we may not need to embed the network,
but use instead the physical coordinates of its nodes to construct the NNG equilibria.
We consider three examples: the
Hungarian road network, the airport network of the United
States, and a structural network of the human brain.
In the first network the nodes are the cities, towns, and villages of
Hungary, while in the second network the nodes are US airports. Two nodes
are linked if they are connected by a direct road or flight.
In the brain network the nodes are small regions of average size
$1.5\mathrm{cm}^2$ covering entirely both hemispheres of the cerebral cortex,
and two regions are connected if a structural connection between them is detected
in diffusion spectrum imaging.
We expect the NNG to be particularly accurate in predicting links in these
networks using the physical---instead of
hyperbolic---coordinates of nodes.
We note that these physical coordinates are Euclidean in all the three cases. The
embedding space is two-dimensional Euclidean and spherical space in the road and
airport cases, and it is three-dimensional Euclidean space in the brain case.
Our method to construct an NNG equilibrium applies without change to
any set of points in any geometric space, and the analytic results on the structure of NNG equilibrium networks in Euclidean spaces are in
Appendix~11. We apply our method to find the NNG equilibrium networks using
the physical coordinates of nodes in these three real networks, and then compare them to their NNG equilibria
also in Figure~\ref{fig:5} and Table~\ref{tab:tab1}.

We observe that in the brain and road networks the NNG link prediction accuracy
is particularly high, reaching $89\%$ for all the links and
$94$-$95\%$ for the frame links. For the brain this result implies
that the spatial organisation of the brain is nearly
optimal for information transfer, in agreement with previous results~\cite{Tiesinga2001Optimal,Laughlin2003Communication,Heuvel2012Brain,Goni2014Restin-brain}.
In the Hungarian road network, nearly all frame
links, crucial for efficient navigation using geography,
are present.
Practically this means that Hungarians have luxury to go on a road
trip without a map since all the major roads required by geographic
navigation are there, albeit the condition of some of those roads
is not as luxurious. To put it simply, there are roads where
people with a compass may think they should be.

For the US airport network however, the geographic results are
poor. These poor results may be unexpected at first, but they have a
simple explanation in that the geometry of the airport network is not
really Euclidean, as the geometry of the nearly planar road network,
but hyperbolic. Indeed, efficient paths in the airport network
optimise not so much the geographic distance travelled, but the number
of connecting flights. As a consequence, most paths go via hubs. As
opposed to the road network, where the number of roads meeting at an
intersection does not vary that much from one intersection to another,
the presence of hubs in the airport network makes the network
heterogeneous, i.e., node degrees vary widely. This heterogeneity
effectively creates an additional dimension (the ``popularity''
dimension in~\cite{PaBoKr11}). That is, in addition to their
geographic location, airports also have another important
characteristic---the size or degree. This extra dimension makes
the network hyperbolic~\cite{KrPa10}. The NNG results for
the hyperbolic map of the airport network in Figure~\ref{fig:5} are as
good as for the other networks.

\noindent \textbf{How to cure or injure a network efficiently.}

The knowledge of the NNG equilibrium of a given real network makes it
possible to efficiently identify links that are most critical for
navigation in the network.  Since NNG equilibrium networks are
maximally navigable networks composed of the smallest number of links,
we expect that if we alter a real network by either adding or removing
a relatively small number of links belonging to the NNG equilibrium of
the network, then such network modifications may significantly affect
network navigability.

\begin{figure}[H]
  \centering
  \includegraphics[width=0.45\textwidth,angle=0]{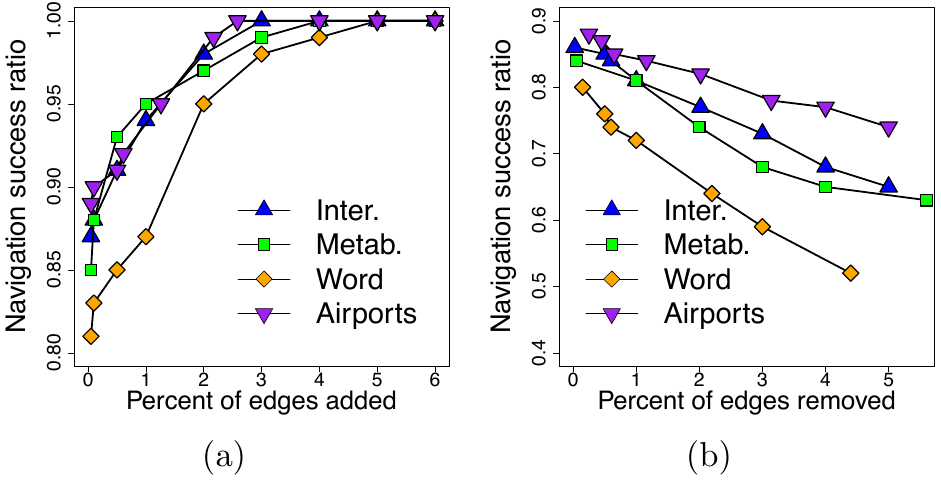}
  \caption{NNG equilibria of real networks helps to improve or degrade
    their navigability. The edges from the NNG equilibria of the
    considered real networks are first sorted in the decreasing order
    of betweenness centrality, and then either added to the real
    network if not already there (panel (a)), or removed from the
    network if present (panel (b)). The $x$-axis shows the percentage
    of added or removed edges compared to the number of edges in the
    original real network. Navigation success ratio is computed as the
    number of node pairs between which geometric routing is successful
    divided by the number of all node pairs. \label{fig:ncorr}}
\end{figure}

Figure~\ref{fig:ncorr} supports these expectations. In the figure, we
take the considered real networks, and add to them certain numbers of
links that are present in the NNG equilibria of the real networks, but
not present in the networks themselves. About 1-2\% of added edges,
compared to the original numbers of edges in the networks, increase
network navigability significantly, while the addition of 2-5\% of
edges makes all the networks 100\%-navigable.  Similarly, the targeted
removal of a small portion (1-5\%) of edges belonging both to the NNG
equilibria of the networks, and to the network themselves, degrades
network navigability by 10-30\%.

\section{Discussion}

We emphasise that the considered Nash equilibrium networks are
minimalistic idealisations, concerned only with maximising the
efficiency of the navigation function at minimal cost (number of
links). Reality differs from this ideal in many ways. First, real
networks must be robust with respect to noise and random
failures. This robustness requirement explains why the considered real
networks have strictly more links that their Nash equilibria. Maximum
navigability can obviously be achieved not only at the minimal cost,
but also at a higher cost. Second, transport processes in real
networks are also noisy, and can follow not only steepest descent path
(greedy navigation), but also any downstream paths, still achieving
$100\%$ reachability. Yet the noisier the transport process, the less
likely it stays to the shortest path, leading to higher stretch and
longer travel times, thus degrading navigation efficiency in terms of
these parameters. Third, navigability does not always have to be
maximised as many specific networks perform many specific functions
other than navigation. Our game-theoretic approach can be extended to
accommodate some of these functions, such as error tolerance or policy
compliance~\cite{szabo2013notes}, but not all possible functions of
different real networks can be formalised within this game-theoretic
framework.  Some networks are centrally designed to optimise a
particular function globally~\cite{lee2013greedy}. Game theory is not
needed to formalise such global optimisation strategies.  It is more
suited for self-organised networks, in which each node behave
selfishly according to its own incentive, independent of other
nodes. In other words, Nash equilibrium networks are structural
manifestations of local incentives of nodes for efficient transport or
communication, in contrast with existing generative or optimisation
models of complex networks~\cite{d2007emergence,Muchnik2013Origins}.
Finally, all real networks are dynamic and growing, while Nash
equilibria correspond to static network configurations. However it has
been recently shown~\cite{KrOs13} that in case of random geometric
graphs---to which the considered Nash equilibrium networks effectively
belong according to the results in Appendix~12---one can map
an equilibrium network model to an identical growing one.

Notwithstanding these limitations we have shown that ideal networks
designed to be maximally navigable at minimal cost share basic
structural properties with real networks. Compared to existing works
on navigation-optimal distributions of shortcut edges in Euclidean
grids
\cite{kleinberg00-nature,li2010towards,rozenfeld2010small,li2013optimal}
which do not yield realistic network topologies, this result is quite
unexpected because there is absolutely nothing in the definition of
these ideal networks that would enforce or even welcome a formation of
any particular network structure. The networks are defined purely in
terms of navigation optimality. The surprising finding that the
structure of these ideal networks is similar to the structure of real
networks should not be misinterpreted as if these idealisations are
generative models for real navigable networks. Instead the former are
skeletons or subgraphs of the latter. Since these skeletons consist of
the minimum number edges required for $100\%$ navigability,
there is no even a parameter to control the most basic structural
network property---the average degree, which is always controllable in
generative models. On the contrary, as follows from
Eq.~(\ref{eq:ave-deg}), the average degree in these skeletons is
uncontrollable and lies between~$1$ and~$4$.

We find that if network geometry is hyperbolic, then our navigation
skeletons have power-law degree distributions and strong clustering.
The values of power-law exponent~$\gamma$ close to~$2$, observed in
many real networks~\cite{newman03c-review,BoLaMoChHw06}, appear as the
best possible choice. In this case not only reachability is $100\%$,
but also the network cost and stretch are minimised and navigability
robustness is maximised, compared to other values of~$\gamma$ in
Figure~\ref{fig:4}.

These results apply to sets of points in hyperbolic space, but the
navigation skeleton construction itself is by no means limited to
these hyperbolic settings. It is very general, and applies to any set
of points in any geometric space, as illustrated by the brain and road
networks where we have used the Euclidean $2d$ and $3d$ physical
coordinates of nodes to construct the navigation skeleton of the
network. Our finding that the brain contains almost fully its
navigation skeleton appears as a mathematically clear and conclusive
evidence that the spatial organisation of the brain is nearly optimal
for communication and information transfer, corroborating existing
work on the
subject~\cite{Tiesinga2001Optimal,Laughlin2003Communication,Heuvel2012Brain,Goni2014Restin-brain}.

We note that the connection between the structure and function of
networks is often studied in the logically reverse direction:
structure$\to$function. That is, first some data about the structure
of real networks is obtained, and then questions concerning how
optimal this structure is with respect to a given network function are
investigated. This logic does provide some evidence that the network
might have evolved optimising this function, but this evidence is
quite indirect and unreliable compared to the direct demonstration
that functionally optimal networks have the structure observed in
reality: function$\to$structure. The common sense suggests that this
causal direction must reflect reality more adequately since networks,
either designed or naturally evolving, do not have a completely random
structure but the structure (effectively) optimising some
functions. Yet studying networks in this direction is much more
challenging primarily because of difficulties in formalising the
constraints that a given function imposes, and deriving the resulting
optimal network structure. Here, with the help of game theory, we have
done so for the navigation function that many real networks
(implicitly) perform.

As one would logically expect, the function$\to$structure approach
provides a deeper insight into specific details of network's
structural organisation that are critical for its functional
efficiency. We have confirmed this expectation by demonstrating that
our approach can identify links in real networks that are most
critical for navigation. A targeted attack on these critical links
degrades navigability rapidly, while if a real network is not
$100\%$-navigable, our approach finds the minimal number of
not-yet-existing links whose addition to the network boosts up its
navigability to $100\%$. Therefore our approach can be used to
identify real network links that should be protected most in a
critical network infrastructure. On the other hand, this approach can
also help network designers to prioritise possible link placement
options, i.e., pairs of not directly connected nodes, that, if
connected, would maximise navigability improvement.

Finally, all the real networks considered here are expected to be
navigable. Indeed, the primary functions of the Internet, brain,
metabolic, or airport and road networks are to transport information,
energy, or people. Semantic and syntactic navigability of word
networks is an established fact in cognitive
science~\cite{cancho2001smallworld,choudhury2009structure,baron2013networks}.
However one cannot expect all real networks to be highly navigable as
navigation is not an important function of every network in the world.
In Appendix~13 we consider one example, a technosocial web
of trust, in which nodes are public keys of users of a distributed
cryptosystem, linked by users' certifications of key-user bindings.
There is no reason why this network should be navigable. In agreement
with this observation, we then find that this network does not contain
a large percentage of edges from its NNG equilibrium, suggesting that
the introduced methodology can be also used as a litmus test to
investigate if navigation is an important function of a given real
network, and if so, then to what degree.

\section{Methods}

\noindent \textbf{The real network data.}  The Internet dataset
representing the global Internet structure at the Autonomous System
(AS) level is from~\cite{BoPa10}.  The metabolic network is the
post-processed network of metabolic reactions in {\it E.~coli}
from~\cite{PaBoKr11}, Snapshot $S_1$ there. The post-processing
details can be found in~\cite{PaBoKr11}. The word network is the
largest connected component of the network of adjacent words in
Charles Darwin's ``The Origin of Species''
from~\cite{MiItKaLeSoAyShAl04}.  The airport network was downloaded
from the Bureau of Transportation Statistics
\url{http://transtats.bts.gov/} on November 5, 2011.  The structural
human brain network and physical coordinates of nodes (regions of
interest (ROIs)) in it are the diffusion spectrum imaging (DSI) data
from~\cite{Hagmann2008DSI}.

\noindent \textbf{The hyperbolic maps of real networks.}
The hyperbolic coordinates of ASes and metabolites are from~\cite{BoPa10}
and~\cite{PaBoKr11}.
The hyperbolic coordinates of words and airports
are inferred using the {\it HyperMap} algorithm~\cite{PaPs14}.
This algorithm is deterministic, and is based on
the growing network model in~\cite{PaBoKr11} used to show that
the latent geometry of scale-free strongly clustered real networks
is hyperbolic. Given an adjacency matrix of a real network, the
algorithm infers the hyperbolic coordinates of its nodes by replaying
its growth as the model in~\cite{PaBoKr11} prescribes. Specifically,
the nodes are first sorted in the order of decreasing degrees, and then,
starting with the highest-degree node, nodes and their edges are added, one node at a time,
to a growing network. The probability, or the likelihood, with which model~\cite{PaBoKr11}
generates this growing network, depends on the node coordinates. The coordinate
of each added node is set by the HyperMap algorithm to the coordinate
corresponding to the global maximum of this probability.

\noindent \textbf{The Nash equilibrium networks of NNGs.}  The
hyperbolic or physical, in the airport and brain cases, coordinates
are then supplied to the GNU Linear Programming Kit (GLPK)
\url{http://www.gnu.org/software/glpk/} used to find a solution to the
corresponding minimum set cover problem.  To yield acceptable running
times of the solver, the Internet and word networks are reduced in
size by extracting their high-degree cores of about 4500 nodes.  The
Hungarian road data is processed slightly differently. First the
cities in Hungary are mapped to their geographic coordinates using the
database in
\url{http://www.kemitenpet.hu/letoltes/tables.helyseg_hu.xls}.  Then
these coordinates are used in the GLPK to find the NNG
equilibrium. Each edge in this equilibrium network is then checked for
existence in the real road network. To check that, the GoogleMaps API
\url{https://pypi.python.org/pypi/googlemaps/} is used to find the
shortest path between the two cities connected by the edge. The edge
is defined to also exist in the real road network if this shortest
path does not go via any other city.

\section{Acknowledgements} We thank Olaf Sporns for sharing the brain
  data, and János Tapolcai and Alexandra Aranovich for useful
  discussions and suggestions.  This work was supported by DARPA grant
  No.\ HR0011-12-1-0012; NSF grants No.\ CNS-1344289, CNS-1442999,
  CNS-0964236, CNS-1441828, CNS-1039646, CNS-1345286, CCF-1212778 and
  Hungarian Scientific Research Fund (grant No. OTKA 185101); and by
  Cisco Systems.  G.R. was supported by the OTKA/PD-104939 grant and
  J.J.B. was supported by the Inter University Centre for
  Telecommunications and Informatics (ETIK), Hungary.

\setcounter{equation}{0}

\graphicspath{{./SI/}}
\begin{centering}
  \section{Appendix 1 - Formal definition of the Network Navigation Game (NNG)}
\end{centering}

\textbf{Strategies.} The strategy space for a player $u\in \mathcal{P}$
is to create some set of arcs to other players in the network: $S_u =
2^{\mathcal{P}\setminus\{u\}}$. Let $s$ be a strategy vector: $s = (s_0, s_1
\dots s_{N-1}) \in (S_0, S_1 \dots S_{N-1})$ and $G(s)$ be the graph defined by
the strategy vector $s$ as $G(s)=\bigcup_{i=0}^{N-1}(i \times s_i)$.

\textbf{Payoff.}
The objective of the players is to minimise their
cost which is calculated as:
\begin{equation}
\label{eqn:gnfgcost}
c_u=\sum_{ \forall u \ne v} d_{G(s)}(u,v)+|s_u|, \quad u,v \in \mathcal{P}
%c_u=\underbrace{\frac{\alpha}{N}\sum_{ \forall u \ne v} d_{G(s)}(u,v)}_{\textrm{communication costs}}+\underbrace{(1-\alpha) |s_u|}_{\textrm{link costs}}, \quad u,v \in \mathcal{P}
\end{equation}
where
\begin{equation*}
d_{G(s)}(u,v)
=\left\{
\begin{array}{cc}
0 & \exists \text { } u \rightarrow v \text{ greedy path in } G(s)  \\
\infty & \text{otherwise.} \\
\end{array}%
\right.
\end{equation*}\\

\begin{centering}
  \section{Appendix 2 - NNG equilibrium}
\end{centering}
%\subsection{Formal calculation of the Nash equilibrium}

The Nash equilibria of the Network Navigation Game can be characterised for each player
independently as follows: take a player
$u$, and for all $v \in V\setminus{u}$ let $\mathbb{S}_v^u=\{w|
d(v,w)<d(u,w)\}$. Trivially $\mathbb{S}_v^u \subset V$ and $\bigcup_{v \in
V\setminus{u}} \mathbb{S}_v^u = V$. The optimal strategy $s_u^{\text{opt}}$ of
$u$ is the minimal set cover of $V$ with the sets $\mathbb{S}_v^u$,
independently from the strategies of the other players. This means that
$s=(s_1^{\text{opt}},s_2^{\text{opt}}\dots s_{N-1}^{\text{opt}})$ is both a NE
and a social optimum.

The Nash equilibrium is not necessarily unique as there can exist different
solutions of the above set cover problem. In our work we concentrate on a specific equilibrium, which besides being a solution, it also minimises the sum of edge the edge lengths all over the network. This is fully in line with the edge-locality principle of complex networks \cite{watts2002identity} \cite{kleinberg2000navigation} \cite{boguna2009navigability} which many times accounted for the high clustering coefficient. More formally, from the strategy vectors constituting a Nash equilibria $s_i$ and the corresponding graphs $G(s_i)=\bigcup_{i=0}^{N-1}(i \times s_i)=(V,E_i)$ we seek for the one minimising $\sum_{j\in E_i}d(E_i(j))$.\\

\subsection{An example}

%\subsection{A small Example in the Euclidean plane}

\begin{figure}[h]
\begin{center}
    \includegraphics[width=.3\textwidth,angle=0]{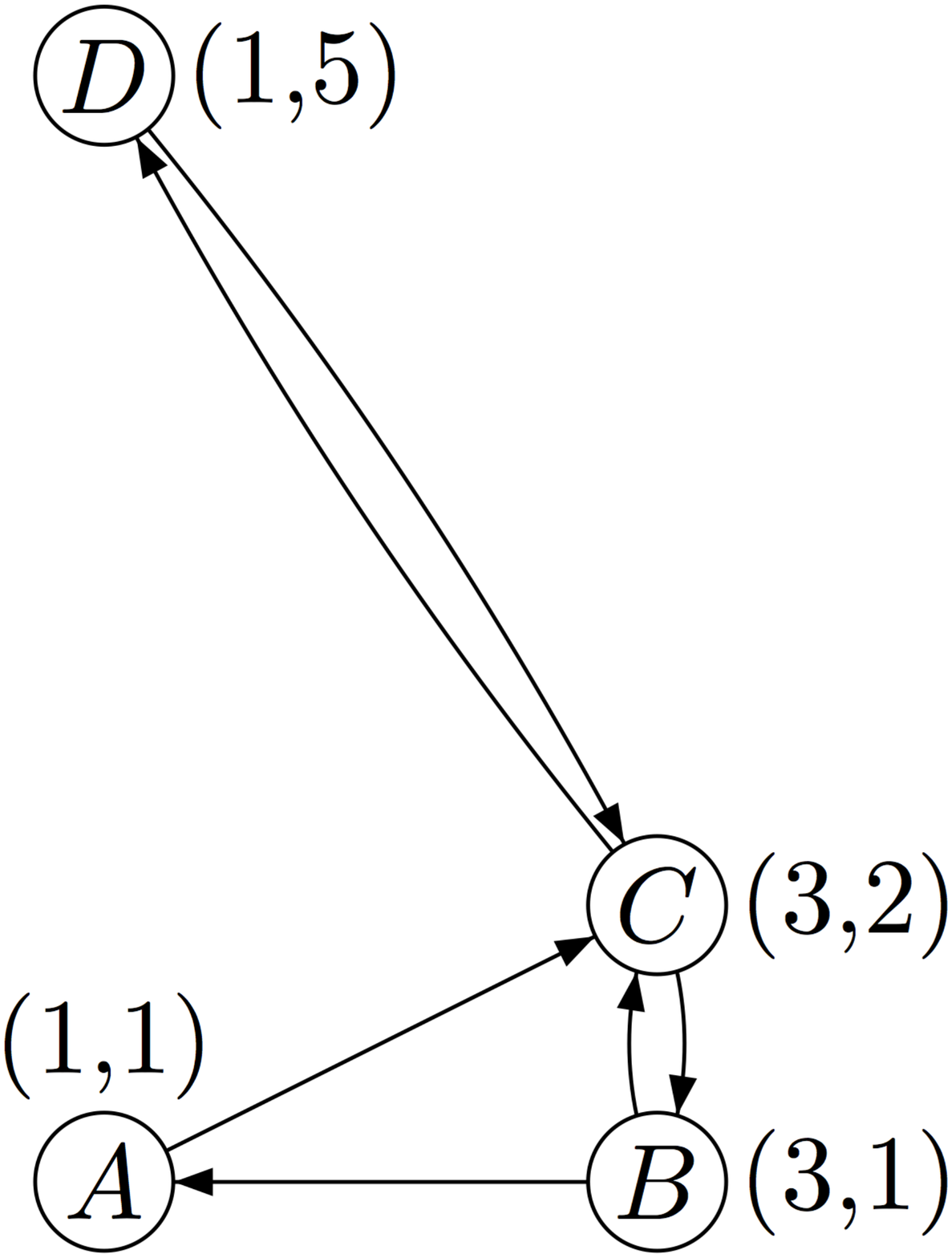}
    \caption{Network Navigation Game in the Euclidean plane.}
    \label{fig:greedy_vs_shortest}
\end{center}
\end{figure}

As an example, let us compute the Nash equilibrium topology for four
points in the Euclidean plane $A,B,C,D$ (see
Figure~\ref{fig:greedy_vs_shortest}).
Any node $u$ out of these four
needs to have a greedy next hop towards any other nodes (to avoid
infinite cost), while having its number of edges minimised. Note that
having a greedy next hop is sufficient since all the other nodes will
have greedy next hops towards any other nodes for ensuring $c_u\le
\infty$. This will imply greedy paths between arbitrary pairs of
nodes.

Let us compute the sets $\mathbb{S}_v^u=\{w|d(v,w)<d(u,w)\}$ for the nodes, where $d(x,y)$ is the Euclidean distance and the minimal set covers for each node to get the Nash equilibrium.
\begin{itemize}
\item $\mathbb{S}_A^B=\{w|d(A,w)<d(B,w)\}=\{A,D\}$, which means that $A$ is a good greedy next hop towards $A$ and $D$ for $B$, similarly $\mathbb{S}_C^B=\{C,D\}$, $\mathbb{S}_D^B=\{D\}$ therefore the minimal cover for $B$ is \{A,C\} so $B$ creates two edges to $A$ and $C$
\item $\mathbb{S}_B^A=\{B,C\}$, $\mathbb{S}_C^A=\{C,B,D\}$, $\mathbb{S}_D^A=\{D\}$ therefore the minimal cover for $A$ is \{C\} so $A$ creates one edge to $C$
\item $\mathbb{S}_A^C=\{A\}$, $\mathbb{S}_B^C=\{B,A\}$, $\mathbb{S}_D^C=\{D\}$ therefore the minimal cover for $C$ is \{B,D\} so $C$ creates two edges th $B$ and $D$
\item $\mathbb{S}_A^D=\{A,B,C\}$, $\mathbb{S}_B^D=\{B,C,A\}$, $\mathbb{S}_C^D=\{C,B,A\}$ therefore the minimal cover for $D$ is  for example \{C\}  ($A$ and $B$ would be also good) so $D$ creates one edge to $C$
\end{itemize}
Thus we can construct the graph from these minimal set coverings see Figure~\ref{fig:greedy_vs_shortest}.
This is a Nash equilibrium and a social optimum as there are no lower
cost equilibria or state for this game.

\begin{centering}
  \section{Appendix 3 - Frame topology}
\end{centering}

%\section{The frame topology}

There exists a well defined ``frame topology'' $G_{\mathrm{frame}}$
with scale-free out-degree distribution which is present in {\em
  every} Nash equilibrium, or social optimum of the NNG
($G_{\mathrm{frame}} \subset G(s^*)$) and other possible games having
navigation as an incentive ($p_s=1$). In other words the frame
topology serves as a skeleton of any equilibrium topology emerging
from navigational games.  The frame topology is defined as:
\begin{Def}[Frame topology]
Let $G_{\mathrm{frame}}=\bigcup_{u=0}^{N-1}(u \times g_u) $, where $g_u=\{v| v\notin s_u \Rightarrow c_u=\infty\}$.
\end{Def}

\begin{figure}[h]
 \label{frame_edge}
\begin{center}
\begin{tikzpicture}[scale=2,cap=round,
angle/.style={font=\fontsize{7}{7}\color{black}\ttfamily} ]

%begin drawings

\coordinate (O2) at (2.0,1.5); %center of the right disk
\coordinate (B2) at ($ (O2) + (0.7,0) $); \coordinate (B3) at ($ (O2) +
(0.2,0.4)$);

\draw [ name path = D2] (O2) circle (0.9); \node (C2) at (B2) [draw, circle
through=(B3)] {};

\begin{scope}
 \clip (B2) circle (0.6403);
 \fill[color=gray!30,shade] (O2) circle (0.9);
\end{scope}

\draw [ name path = D2 ] (O2) circle (0.9); \node (C2) at (B2) [draw, circle
through=(B3)] {};

\node [left] at (O2) {$O$}; \node [right] at (B2) {$v$}; \node [left] at (B3)
{$u$}; \node [left] at (B2) {$T_{u,v}$};

\foreach \point in {O2,B2,B3}
    \fill [black,opacity=.5] (\point) circle (0.5pt);

\draw [arrows={-latex}] (B3) -- (B2);

 \end{tikzpicture}
\caption{An edge in the $G_{\mathrm{frame}}$} \label{fig:frame_edge}
\end{center}
\end{figure}
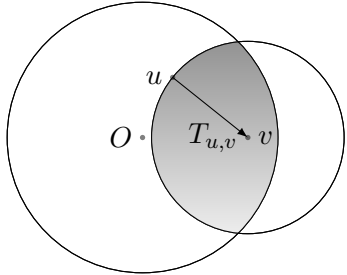

Practically, the arc $(u, v)$ is contained in $G_{\mathrm{frame}}$ if and only if the
$d(u,v)$-disk centred at $v$ does not contain any player other than $u$ (see
Figure \ref{fig:frame_edge}). 
This means that $u$ cannot reach $v$ by greedy
routing through any other players then $v$, and so it must create an arc
towards $v$ to avoid of having infinite cost. Note that the in-degree of each
player in $G_{\mathrm{frame}}$ will be exactly one.

%\begin{thm}
\begin{centering}
\section{Appendix 4 - Connection probability}
\end{centering}

Here we cast the problem in statistical terms. We estimate the percentage of pairs of nodes located at a given distance that are connected in the NNG equilibrium. We call this percentage the effective connection probability.
First the connection probability of the Frame Topology is derived.  This connection probability is a lower bound for the connection probability in the NNG equilibrium network because the Frame Topology is contained in every NNG equilibrium network. A direct upper bound of the connection probability is also studied. Based on a statistically equivalent lower bound and the direct upper bound, a general formula for the connection probability is induced, in which the average degree of the network is implicitly encoded. This makes it possible to approximate the connection probability (and all other quantities defined by it) using the observed average degree in the NNG simulation.\\

\subsubsection{Connection probability in the Frame Topology}

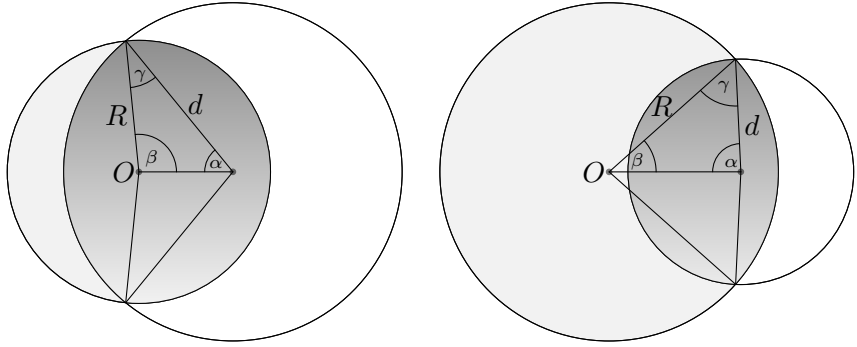
\begin{figure}[htb]
\begin{center}
\begin{tikzpicture}[scale=2.5,cap=round,
angle/.style={font=\fontsize{7}{7}\color{black}\ttfamily} ]

%begin drawings

\coordinate (O2) at (2.0,1.5); %center of the right disk
\coordinate (B2) at ($ (O2) + (0.7,0) $);
\coordinate (O1) at (-0.5,1.5); %center of the left disk
\coordinate (B1) at ($ (O1) + (0.5,0) $);

\draw [ name path = D2,fill=gray!10 ] (O2) circle (0.9); \draw [ name path = C2
] (B2) circle (0.6); \draw [ name path = D1,fill=gray!10] (O1) circle (0.7);
\draw [ name path = C1 ] (B1) circle (0.9);

\begin{scope}
\clip (B2) circle (0.6); \fill[color=gray!30,shade] (O2) circle (0.9);
\end{scope}

\begin{scope}
\clip (O1) circle (0.7); \fill[color=gray!30,shade] (B1) circle (0.9);
\end{scope}

\draw [ name path = D2 ] (O2) circle (0.9); \draw [ name path = C2 ] (B2)
circle (0.6); \draw [ name path = D1 ] (O1) circle (0.7); \draw [ name path =
C1 ] (B1) circle (0.9);

\node at ($(O1)-(0.08,0)$) {$O$}; \node at ($(O2)-(0.08,0)$) {$O$}; \foreach
\point in {O1,O2,B1,B2}
   \fill [black,opacity=.5] (\point) circle (0.5pt);

%right part
\path [name intersections={of=D2 and C2, by={A,B}}]; \draw (O2) -- (A); \draw
(O2) -- (B); \draw (B2) -- (A); \draw (B2) -- (B); \draw (B2) -- (O2);

% we may subtract 360 from the \AngleEnd values

\tikzAngleOfLine(O2)(A){\AngleStart} \tikzAngleOfLine(O2)(B2){\AngleEnd} \draw
(O2)+(\AngleStart:0.25) arc (\AngleStart:\AngleEnd:0.25); \node [angle] at
($(O2)+({(\AngleStart+\AngleEnd)/2}:0.16)$) {$\beta$};

\tikzAngleOfLine(A)(O2){\AngleStart} \tikzAngleOfLine(A)(B2){\AngleEnd} \draw
(A)+(\AngleStart:0.25) arc (\AngleStart:\AngleEnd:0.25); \node [angle] at
($(A)+({(\AngleStart+\AngleEnd)/2}:0.16)$) {$\gamma$};

\tikzAngleOfLine(B2)(O2){\AngleStart} \tikzAngleOfLine(B2)(A){\AngleEnd} \draw
(B2)+(\AngleStart:0.15) arc (\AngleStart:\AngleEnd:0.15); \node [angle] at
($(B2)+({(\AngleStart+\AngleEnd)/2}:0.07)$) {$\alpha$};

\draw ($ (O2) + (0.28,0.35) $) node {$R$}; \draw ($ (O2) + (0.76,0.25) $) node
{$d$};

%left part
\path [name intersections={of=D1 and C1, by={A,B}}]; \draw (O1) -- (A); \draw
(O1) -- (B); \draw (B1) -- (A); \draw (B1) -- (B); \draw (B1) -- (O1);

% we may subtract 360 from the \AngleEnd values
\tikzAngleOfLine(O1)(A){\AngleStart} \tikzAngleOfLine(O1)(B1){\AngleEnd} \draw
(O1)+(\AngleStart:0.2) arc (\AngleStart:\AngleEnd:0.2); \node [angle] at
($(O1)+({(\AngleStart+\AngleEnd)/2}:0.1)$) {$\beta$};

\tikzAngleOfLine(A)(O1){\AngleStart} \tikzAngleOfLine(A)(B1){\AngleEnd} \draw
(A)+(\AngleStart:0.25) arc (\AngleStart:\AngleEnd:0.25); \node [angle] at
($(A)+({(\AngleStart+\AngleEnd)/2}:0.2)$) {$\gamma$};

\tikzAngleOfLine(B1)(O1){\AngleStart} \tikzAngleOfLine(B1)(A){\AngleEnd} \draw
(B1)+(\AngleStart:0.15) arc (\AngleStart:\AngleEnd:0.15); \node [angle] at
($(B1)+({(\AngleStart+\AngleEnd)/2}:0.10)$) {$\alpha$};

\draw ($ (O1) + (-0.12,0.30) $) node {$R$}; \draw ($ (O1) + (0.30,0.36) $) node
{$d$};

\end{tikzpicture}
\end{center}
\caption{Illustration for $T_{uv}$ \label{fig_Tuv}}
\end{figure}

As presented in Appendix 3 an arc $(u,v)$ in the Frame Topology is established if and only if there are no other points (players) within the intersection of the $v-$centred disk with radius $d(u,v)$, and the original disk with radius $R$.
The probability of this event is
\begin{equation}
\left(\frac{T_R - T_{uv}}{T_R}\right)^{N-2} \approx e^{-\delta T_{uv}}
\end{equation}
An approximation for $T_{uv}$ is as follows: $T_{uv}$ is apparently
equals to $2 \pi (\cosh d_{uv}-1) (\approx \pi e^{\frac{d_{uv}}{2}}$ for not so
small $d_{uv}$) when the $d_{uv}-$disk is completely inside the $R-$disk.
On contrary, if $R-r_v < d(u,v)$ (there is real intersection) then much less
evidently $T_{uv}$ is approximately $T_{uv} \approx 4 e^{\frac{d_{uv}}{2}}
e^{\frac{R-r_v}{2}}$.  In
Figure \ref{fig_Tuv} two characteristic cases are depicted when there is real
intersection of the $d_{uv}$-disk and the $R$-disk. 
Let the polar coordinates
of node $v$ be $(r_v,\phi_v)$, and of node $u$ be $(r_u, \phi_u)$. Let $\phi =
|\phi_u - \phi_v|$. The area $T_{uv}$ is the function of $r_u$, $r_v$, $\phi$,
and $R$, and can be calculated as
the sum of the two circle sectors with angle $2 \alpha$, radius $d_{uv}$ and
angle $2 \beta$ radius $R$, and minus the area of the two triangles with angles
$\alpha, \beta,\gamma$. That is
\begin{equation} \label{equ:Tuv}
T_{uv} = 2\beta \left( \cosh(R)-1\right) + 2\alpha \left(
\cosh(d_{uv})-1\right) -2 \left( \pi -\alpha -\beta-\gamma \right) \ .
\end{equation}
where the angles and $d_{uv}$ are given by the
hyperbolic law of cosines, however, here the following simpler approximations
are used (which are accurate enough when $r_u$ and $d_{uv}$ appear in exponents):
\begin{equation} \label{equ:duvapprox}
d_{uv} \approx R + r_v +2 \ln \frac{\beta}{2} \Rightarrow \beta \approx 2
e^{\frac{d_{uv}}{2} - \frac{R+r_v}{2}}
\end{equation}
\begin{equation}
R \approx d_{uv} + r_v +2 \ln \frac{\alpha}{2} \Rightarrow \alpha \approx 2
e^{-\frac{d_{uv}}{2} + \frac{R-r_v}{2}} \ .
\end{equation}
Applying (\ref{equ:Tuv}) with neglecting the triangle areas, and using
$\cosh(R)-1 \approx e^{R}/2$, $\cosh(d_{uv})-1 \approx \frac{e^{d_{uv}}}{2}$ we
get $T_{uv} \approx 4 e^{\frac{d_{uv}}{2}} e^{\frac{R-r_v}{2}}$.

\begin{figure}[htb]
\begin{center}
  \includegraphics[width=.9\textwidth,angle=0]{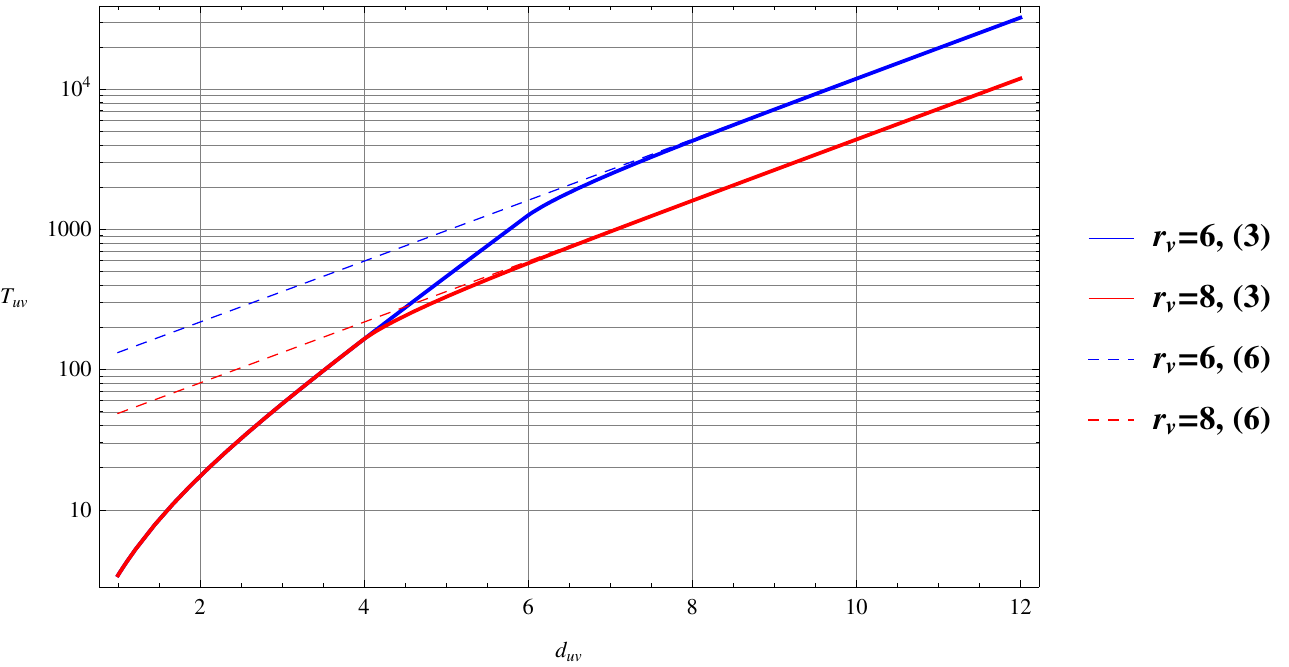}
     \caption{$T_{uv} \approx 4 e^{\frac{d_{uv}}{2}} e^{\frac{R-r_v}{2}}$ when there are real intersections (that is when $d(u,v) > 6$ and $4$,
     respectively).}
    \label{Fig_Tuv_proportional}
\end{center}
\end{figure}

In summary:
\begin{equation} \label{equ:Tuv_summary}
T_{uv} \approx \left\{
\begin{aligned}
\pi e^{d_{uv}}                                  && \textrm{, if} && 0 < d_{uv} < R-r_v \\
4 e^{\frac{d_{uv}}{2}} e^{\frac{R-r_v}{2}}      && \textrm{, if} && d_{uv} >
R-r_v \ .
\end{aligned}
 \right.
\end{equation}
This $T_{uv}$ approximations are illustrated in
Figure~\ref{Fig_Tuv_proportional} for $R=12$, $r_v=6$ and $r_v=8$.
Solid lines are the exact $T_{uv}$ calculations
based on (\ref{equ:Tuv}) and exact computations of angles. Note that there is a
sharp change on logarithmic scale between the $d_{uv}$-slope and $d_{uv}/2$-slope
around $R-r_v$. The dashed lines are the $T_{uv}$ approximations when $d_{uv} >
R-r_v$.
%Appendix Figure~\ref{fig:rel_error_rv6} and
%Appendix Figure~\ref{fig:rel_error_rv8} show the relative error of $1-T_{uv}/(4 e^
%{\frac{d_{uv}}{2}} e^{ \frac{R-r_v}{2}})$ in the possible range of $d_{uv}$: $R-r_v <
%d_{uv} < R+r_v$.
%
%\begin{fiure}[htb]
%\begin{center}
%    \includegraphics[width=.8\textwidth,angle=0]{for_Nature_2013_num_examples_gr2.pdf}
%     \caption{Relative error in the function of $d_{uv}$ in case of $r_v=6$.}
%    \label{fig:rel_error_rv6}
%\end{center}
%\end{figure}
%
%\begin{fiure}[htb]
%\begin{center}
%    \includegraphics[width=.8\textwidth,angle=0]{for_Nature_2013_num_examples_gr1.pdf}
%     \caption{Relative error in the function of $d_{uv}$ in case of $r_v=8$.}
%    \label{fig:rel_error_rv8}
%\end{center}
%\end{figure}
%% todo itt meg kene nezni ponrosabban, hogy van ez a sorfejtes...

The calculation of the expected degree of node $u$ requires $e^{-\delta T_{uv}}$ in the following double integration:
\begin{equation} \label{equ:Tuvforkru}
\delta \int_0^R \int_{0}^{2 \pi} e^{-\delta T_{uv}} {\rm d} \phi \sinh(r_v) {\rm d} r_v \ .
\end{equation}
Because the joint expansion of the double integral with respect to $r_v$ and
$\phi$ reveals that the dominant terms will be those in which $d_{uv} > R - r_v$
\begin{equation}
\delta \int_0^R \int_{0}^{2 \pi} e^{-\delta T_{uv}} {\rm d} \phi \sinh(r_v) {\rm d} r_v \approx
\delta \int_0^R \int_{0}^{2 \pi} e^{-\delta 4 e^{\frac{d_{uv}}{2}}
e^{\frac{R-rv}{2}}} {\rm d} \phi \sinh(r_v) {\rm d} r_v  \ .
\end{equation}
Using (\ref{equ:duvapprox}) it can also be shown that
\begin{equation}
\delta \int_0^R \int_{0}^{2 \pi} e^{-\delta 4 e^{\frac{d_{uv}}{2}}
e^{\frac{R-rv}{2}}} {\rm d} \phi \sinh(r_v) {\rm d} r_v \approx \delta \int_0^R
\int_{0}^{2 \pi} e^{-\delta 8 e^{\frac{d_{uv}}{2}}} {\rm d} \phi \sinh(r_v)
{\rm d} r_v \ .
\end{equation}
Therefore,
\begin{equation}\label{equ:lbcp}
\check p(d_{uv}) = e^{-\delta 8 e^{\frac{d_{uv}}{2}}}
\end{equation}
can be considered as a statistically equivalent connection probability of the Frame Topology and
as a latent (a statistically equivalent) lower bound of the
connection probability of the equilibrium network of the NNG.

\subsubsection{A direct upper bound for the connection probability}

\begin{figure}[h!]
\begin{center}
\begin{tikzpicture}[scale=3.5,cap=round,
angle/.style={font=\fontsize{7}{7}\color{black}\ttfamily}
]
%begin drawings

\coordinate (O) at (0,0); %center of the right disk

\coordinate (R) at (-0.4,-0.8);

\coordinate (U) at (0.3,0.2);
%the point to invert
\coordinate (V) at (0.5,0.6);
\coordinate (C1) at (0.120806, 0.891855);
\coordinate (C2) at (0.805495, 0.40147);

\coordinate (D) at (0.6625, 0.925);
\coordinate (R0) at (0,0.542707);
\coordinate (D1) at (0.0761174, 0.561941);
\coordinate (R1) at (0, 0.425588);
\coordinate (D2) at (0.600953, 0.299523);
\coordinate (R2) at (0, 0.316982);
\coordinate (A) at (0.37, 0.4);

\clip (O) circle (1.0);

\draw[black]  (O) circle (0.9);
\draw[black]  (D1) circle ( 0.425588);
\draw[black] (D2) circle ( 0.316982);
\draw[black]  (D) circle (0.542707);

\begin{scope}
\clip (O) circle (0.9);

\fill[color=gray!30,shade] (D) circle (0.542707);
\end{scope}

\begin{scope}
\clip (D1) circle ( 0.425588);

\fill[color=gray!30,shade] (D2) circle ( 0.316982);
\end{scope}

\node [right] at ($(U)$) {$u$};
\node [below right] at ($(V)$) {$v$};
\node [above left] at ($(V)$) {$C_{uv}$};
\node at ($(A)$) {$A$};
\node [below right] at ($(O)$) {$O$};

\path[draw] (O) -- (R);
\path[draw] (-0.348,0.52) -- (C1);
\path[draw] (0.75,0.02) -- (C2);

\node  at ($(-0.132,0.78)$) {$R_1$};
\node  at ($(0.7,0.17)$) {$R_2$};
\node  at ($(-0.18,-0.47)$) {$R$};
\node  at ($(0.1,0.60)$) {$B_{uv}$};
\fill[black] (0,0) circle (0.01);
\fill[black] (U) circle (0.01);
\fill[black] (V) circle (0.01);
\fill[black] (C1) circle (0.01);
\fill[black] (C2) circle (0.01);
\draw[black]  (O) circle (0.9);
\draw[black]  (D1) circle ( 0.425588);
\draw[black] (D2) circle ( 0.316982);
\draw[black]  (D) circle (0.542707);

\end{tikzpicture}
\caption{Calculation of $p(d_{uv})$ }
\label{fig:ujkrumpli}
\end{center}
\end{figure}
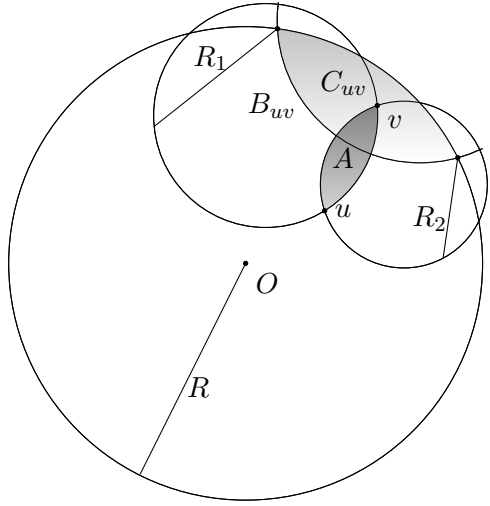

An upper bound for connection probability $p(d_{uv})$ can be derived
as follows. Let $u$ and $v$ be two points in the $R$-disk and let
$C_{u,v}=\{w|d_{wv}<d_{wu}\}$ denote the area for which $v$ is a good
greedy next hop for $u$. This area is on the side of $v$ bounded by
the perpendicular bisector ($B_{uv}$) of $(u,v)$, see
Figure~\ref{fig:ujkrumpli}. (The figure is in the Poincare model).

Let $A=\{x|C_{u,x}\supset C_{u,v}\}$. If there is a node $w\in A$ then $u$ does not connect to $v$ since it has a node $w$ which covers the whole area that $v$ can and some extra portion of the disk. Putting it differently $w$ can be in the optimal set cover (for $u$) instead of $v$.
It is easy to see that $A$ is the intersection of two disks with radii $R_1$ and $R_2$ (the smaller circles on
Figure~\ref{fig:ujkrumpli}). We can approximate the area of this intersection by the union of two sectors having angles $\phi_1\approx 2e^{\frac{d_{uv}}{2}-R_1}$ and $\phi_2\approx2e^{\frac{d_{uv}}{2}-R_2}$ (by using an
approximation on the hyperbolic distance $d_{uv}\approx 2R_i +2
\ln\frac{\phi_i}{2})$ of the $R_1$ and the $R_2$ disks respectively.
Using this the area of $A$ is given by:
\begin{equation}
T_{A}\approx \phi_1 (\cosh(R_1)-1)+\phi_2 (\cosh(R_2)-1)\approx2e^{\frac{d_{uv}}{2}-R_1} \frac{e^{R_1}}{2}+ 2e^{\frac{d_{uv}}{2}-R_2}\frac{e^{R_2}}{2},
\end{equation}
which further simplifies to:
\begin{equation}
T_{A}\approx2e^{\frac{d_{uv}}{2}}.
\end{equation}
The probability that there is a node in $A$ is:
 \begin{equation}
 p(\exists w\in A ) = 1-\left(\frac{T_{\text{disk}}-T_{A}}{T_{\text{disk}}}\right)^{N-2}\approx 1-e^{-\delta T_{A}},
 \end{equation}
 where $N$ denotes the number of nodes and $T_{\text{disk}}$ is the area of the $R$-disk. Trivially $p(d)\le 1-p(\exists w\in A)$ so:
\begin{equation}
p(d_{uv}) \le e^{-\delta T_{A}}.
\end{equation}
By substituting $T_{A}$ we get the following upper bound for the connection probability:
\begin{equation} \label{equ:ubcp}
p(d_{uv})\le e^{-\delta T_{A}} \approx e^{-2\delta e^{\frac{d_{uv}}{2}}} =: \hat p(d_{uv})
\end{equation}

\subsubsection{A general formula for the connection probability}

In the Frame Topology (by definition) every node has exactly one incoming link, hence, the total number of links are $N$. From this it immediately follows that the average out-degree of Frame Topology is 1. This will also confirmed by the results of the next note (Appendix 5), in which the conditional expected degree of a node $u$ with radial coordinate $r_u$ is calculated and shown by un-conditioning that the average degree is 1.
Regarding the direct upper bound of the connection probability, consider a network in which links are established by this upper bound probability. Also the analysis in the next note implies that the average degree of such a network is 4.
Based on the upper (\ref{equ:ubcp}) and lower (\ref{equ:lbcp}) bounds and the corresponding average degrees 1 and 4, a general formula of the connection probability can be induced as
\begin{equation} \label{equ:general_cp}
p(d_{uv},\delta,\bar k) = {\rm Exp}\left(-\frac{8}{\bar k} \delta e^{\frac{d_{uv}}{2}}\right) \ .
\end{equation}
It will be shown in the next sections that a network formed by this connection probability has average degree $\bar k$.

\begin{figure}[h!]
\psfrag{nng}{NNG simulation}
\psfrag{Upp}{Upper bound ($\ref{equ:ubcp}$)}
\psfrag{Lo}{Lower bound ($\ref{equ:lbcp}$)}
\psfrag{App}{App. ($\ref{equ:general_cp}$) $\bar k = 2.27$}
\psfrag{dist}{$d$}
\psfrag{Lnkp}{Link probability $p(d)$}
\begin{center}
    \includegraphics[width=.6\textwidth,angle=0]{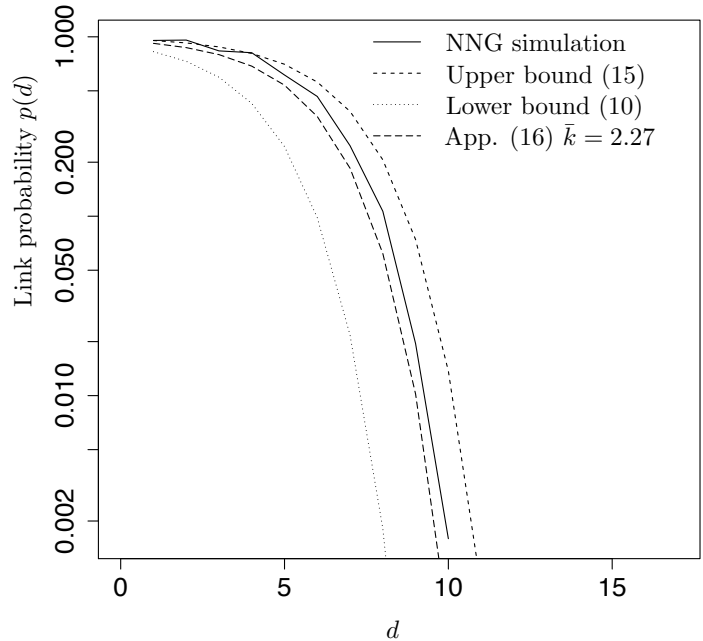}
     \caption{Connection probability as a function of distance in NNG simulations. The figure shows the analytic upper and lower  bounds, as well as the analytic approximation with the empirical mean.}
    \label{fig:connection_probabilities}
\end{center}
\end{figure}

This formula is important because if an empirical average degree (which happens to be $2.27$) can be observed in experiments (simulations) resulting in equilibrium networks of NNG, then not only upper and lower bounds on the expected degree of a node $u$ and degree distribution, but analytical approximations of them can also be given with this empirical mean.
Figure~\ref{fig:connection_probabilities} illustrates the
relation of the upper and lower bounds, and the approximation of the
connection probability to that of simulated NNG.

\begin{centering}
\section{Appendix 5 - Expected degree of a given node}
\end{centering}
The expected (out-)degree of a node $u$ with radial coordinate $r_u$ in a network generated with the effective connection probability formula is given by the following double integral
\begin{equation}
k_{\mathrm{out}}(r_u, \bar k, R) = \delta \int_0^R \int_{0}^{2 \pi} p(d_{uv},\delta, \bar k) {\rm d} \phi \sinh(r_v) {\rm d} r_v \ .
\end{equation}
The expected node-degree of the equilibrium network of NNG is lower bounded by $k_{\mathrm{out}}(r_u,1,R)$ (which coincides the expected node-degree of the Frame Topology) whilst $k_{\mathrm{out}}(r_u,4,R)$ is the upper bound. An analytical approximation with the empirical mean $\bar k =2.27$ can be given by $k_{\mathrm{out}}(r_u,2.27,R)$.

In what follows a formula is derived for $k_{\mathrm{out}}(r_u, \bar k, R)$ based on the integral above.
Considering the first integral by $\phi$
%as
%\begin{equation}
%\delta \int_{0}^{2 \pi} e^ {-\delta T_{uv}} {\rm d} \phi \ .
%\end{equation}
%can be approximated when $T_{uv}$ is substituted by its approximation, that is
%by
%\begin{equation}
%\delta \int_{0}^{2 \pi} e^{-\delta 4 e^{\frac{d_{uv}}{2}}
%e^{\frac{R-rv}{2}}} {\rm d} \phi \ .
%\end{equation}
and applying the approximation (consider the hyperbolic law of cosine for $d_{uv}$, $r_u$, $r_v$, $\cosh d_{uv} = \cosh r_u \cosh r_v - \sinh r_u \sinh r_v \cos \phi$ )
\begin{equation}
e^{\frac{d_{uv}}{2}} \approx
e^{\frac{r_u+r_v}{2}} \sqrt{\frac{1-\cos \phi}{2}}
\end{equation}
%(see equation
%(\ref{equ:ed2approx}))
we get that
%similar form as in (\ref{eqn:besselstruvel}),
%hence
the integral can be approximated as
\begin{equation}
\delta \int_{0}^{2 \pi} {\rm Exp}\left(-\delta \frac{8}{\bar k} e^{\frac{d_{uv}}{2}}\right){\rm d} \phi \approx 2 \pi \delta ({\rm I}(0,x)-{\rm
S}(0,x)) \approx \frac{1}{2} {\bar k} e^{-\frac{r_u+r_v}{2}}
\end{equation}
where $x= \frac{8}{\bar k} \delta e^{\frac{r_u+r_v}{2}}$ and the last wave due to that ${\rm
I}(0,x)-{\rm S}(0,x)$ (difference of the BesselI and the modified Struve
functions) quickly tends to $\frac{2}{\pi} x^{-1}$ as $x$ increases \cite{abramovitz}. Now the second integration
by $r_v$ gives the expected degree approximation, that is
\begin{equation} \label{equ:general_outdegree}%\label{equ:GFkout}
k_{\mathrm{out}}(r_u,\bar k,R) \approx \int_{0}^R \frac{1}{2} {\bar k} e^{-\frac{r_u+r_v}{2}} \sinh(r_v) {\rm
d} r_v \approx \frac{1}{2} {\bar k} e^{\frac{R}{2}}
e^{-\frac{r_u}{2}} \ .
\end{equation}
One can check that the average degree is indeed ${\bar k}$ with this expected node-degree:
\begin{equation}
\int_{r_u=0}^R \frac{1}{2} {\bar k} e^{\frac{R}{2}} e^{-\frac{r_u}{2}} \frac{\sinh r_u}{\cosh R-1} {\rm d} r_u \approx \frac{1}{6} {\bar k}\,\mathrm{sech}^2\left(\frac{R}{4}\right) \left(\sinh \left(\frac{R}{2}\right)+2 \cosh \left(\frac{R}{2}\right)+1\right) \approx {\bar k} \ .
\end{equation}

\begin{figure}[H]
\begin{center}
    \includegraphics[width=.8\textwidth,angle=0]{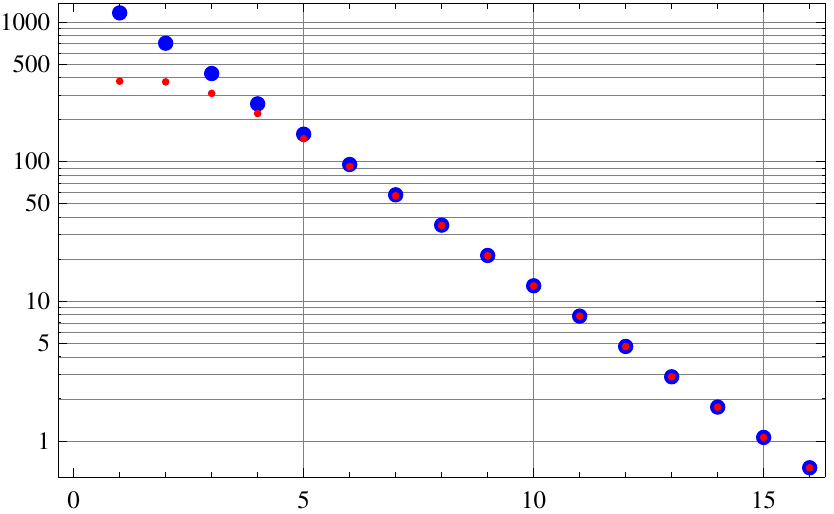}
     \caption{Exponential decay ($C(R) e^{-\frac{ru}{2}}$, larger blue dots)
     versus the numerically evaluated exact decay (smaller red dots) of $u$'s expected degree as a function of $r_u$ ($R=16.5$, $n=10000$).}
    \label{matching_delta_1}
\end{center}
\end{figure}

We have numerically studied the accuracy of the approximations above.
We have found that the exponential decay of the
expected degree of nodes ($k_{\mathrm{out}}(r_u)$) is a good approximation of
the numerically evaluated expected degree function for a wide range of node
density $\delta \in [10^{-8},10^{-2}]$.
For example, consider a Frame Topology (${\bar k}=1$) with
$R=16.5$, $n=10000$. In this case
$\delta=2.17 \cdot 10^{-4}$.
Figure~\ref{matching_delta_1} shows how the expected
degree decay is matching the exponential decay.
We observe that while at smaller $r_u$ there
are some approximation errors, for larger
values of $r_u$ the match is very good.
To quantify further, we note that 99.9\% of points have $r_u > 10$ (that
is in case of uniformly distributed points on the R(=16.5)-disk, expectedly only about 10
points of the 10000 is inside the disk with radius 10). If we consider the
relative errors of the matching one can reveal that for $r_u > 10$ it is
smaller than 0.15\%, that is for 99.9\% of points the expected degree
approximation has smaller than 0.15\% relative error.
%, as shown by
%Appendix Figure~\ref{rel_error_matching}
To increase the number of points to $n=30000$
and $n=50000$ ($\delta = 6.54 \cdot 10^{-4}$, $\delta = 1.08 \cdot 10^{-3}$),
the relative error is increasing, especially for smaller values of $r_u$, but
still for 99.9\% of the points the relative error smaller than 0.25\% and 1\%,
respectively. If we dramatically decrease the node-density, for example
$n=500$, the relative errors also increase (compared to the $n=10000$ case),
however, it still remains under 0.2\% for 99.9\% of the points.\\

%\begin{fiure}[H]
%\begin{center}
%    \includegraphics[width=.8\textwidth,angle=0]{for_Infocom_2013_num_examples_2_gr2.pdf}
%     \caption{Relative error of matching the exponential decay in the function of $r_u$, $R=16.5$, $n=10000$.}
%    \label{rel_error_matching}
%\end{center}
%\end{figure}

\begin{centering}
\section{Appendix 6 - Degree distribution}
\end{centering}

Let us recall that in case of uniform distribution of points on an $R-$disk of the hyperbolic plane, the density of the radial coordinates of the points is
\begin{equation}
\rho(r) = \frac{\sinh r}{\cosh R - 1}
\end{equation}
Note that the expected degree of node $u$ is exponential in the radial coordinate $r_u$ as in \cite{dima_infocom_2010}. Because of this and the fact that equilibrium network of NNG is also sparse \cite{boguna_corr_random_networks} the degree distribution can be calculated in the same way as in \cite{dima_infocom_2010} :
\begin{equation}
P(k) = \int_0^R g(k,k_{\mathrm{out}}(r_u)) \rho(r_u) {\rm d} r_u = \frac{\bar k}{2} \frac{\Gamma(k-2,\frac{\bar k}{2})}{k!}
\end{equation}
where $g(k, k_{\mathrm{out}}(r_u))$ is the conditional distribution of the degree of a node with radial coordinate $u$, and it is Poissonian with mean $k_{\mathrm{out}}(r_u)$ in case of sparse networks.
It can also be shown that for larger $k$
\begin{equation}
P(k) \approx \frac{{\bar k}^2}{2 k^3} \ .
\end{equation}

The direct derivation of the complement cumulant degree distribution from $P(k)$ seems to be intangible, however, from its approximation it can be computed as
\begin{equation}
{\bar F} (k,{\bar k}) \approx 1- \left( \int \frac{{\bar k}^2}{2 k^3} {\rm d} k + C \right)
\end{equation}
where the constant $C$ is 1, and $k \geq \frac{1}{2} {\bar k}$ (in order to have distribution function), that is
\begin{equation}
{\bar F} (k,{\bar k}) \approx \frac{\bar k^2}{4} k^{-2} \ , \ \ k \geq \frac{1}{2} \bar k \ .
\end{equation}

It is interesting to show that this approximation can also be obtained as the \emph{exact} ccdf of the conditional expected node degrees $k_{\mathrm{out}}(r_u)$.
This approximation can be computed as
\begin{equation}
{\bar F} (k,{\bar k}) \approx \int_{r=0}^{r_u(k)} \rho(r) {\rm d} r \approx e^{r_u(k)-R}
\end{equation}
where $r_u(k)$ is the inverse function of $k_{\mathrm{out}}(r_u,{\bar k},R)$ w.r.t. $r_u$, i.e.
\begin{equation}
%r_u(k) = 2 \ln \left( \frac{\bar k}{ 2 k} + R \right) \ .
r_u(k) =  R-2 \ln \left({2 k}/{\bar{k}}\right) \ .
\end{equation}
Applying this one can obtain the same before as
\begin{equation}
{\bar F} (k,{\bar k}) \approx \frac{\bar k^2}{4} k^{-2} \ , \ \ k \geq \frac{1}{2} \bar k \ .
\end{equation}
Note that this yields the average degree equal to ${\bar k}$ as expected:
\begin{equation}
\int_{k=\frac{1}{2} {\bar k}}^{\infty} \left( k \frac{\partial (1-{\bar F}(k,{\bar k}))}{\partial k} \right) = {\bar k}.
\end{equation}

\begin{figure}[h!]
\psfrag{Model}{NNG simulation}
\psfrag{Upp}{Upper bound, $\bar F(k, 4)$}
\psfrag{Lo}{Lower bound, $\bar F(k,1)$}
\psfrag{App}{Approximation, $\bar F(k,2.27)$}
\psfrag{k}{$k$}
\psfrag{F(k)}{$\bar{F}(k,\bar k)$}
\begin{center}
    \includegraphics[width=.7\textwidth,angle=0]{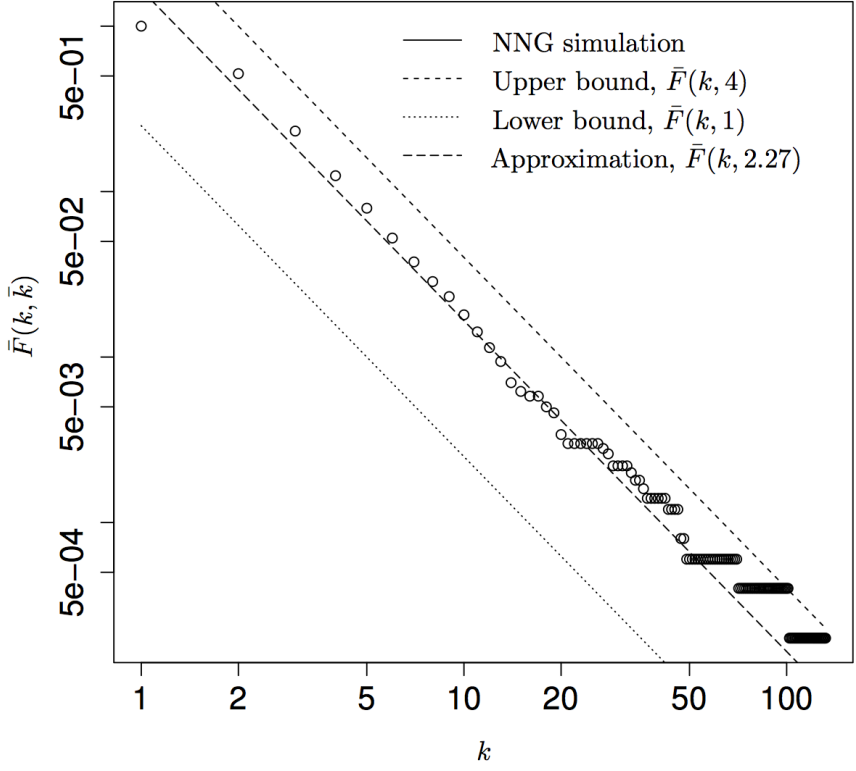}
     \caption{Empirical CCDF of the degree distribution, its analytical upper and lower bounds $\bar F(k,4) ,  \ \bar F(k, 1)$, and analytical approximation with the empirical mean $\bar F(k,2.27)$.}
    \label{fig:dd_anal}
\end{center}
\end{figure}

From this, an analytical approximation of the ccdf of the NNG equilibrium network is ${\bar F} (k,2.27)$, its lower and upper bounds are ${\bar F} (k,1)$, ${\bar F} (k,4)$, respectively. In Figure~\ref{fig:dd_anal} these analytical formulae are drawn also with a completely empirical distribution obtained from NNG simulation.

We also note that the $\delta$-independence of $k_{\mathrm{out}}(r_u)$ and $\bar F(k)$ is approximate, but it holds with a high accuracy for $\delta\in[10^{-8},10^{-2}]$, including the frame topology.\\

\begin{centering}
  \section{Appendix 7 - Clustering}
\end{centering}

Here we analyse local clustering
using the effective connection probability~(\ref{equ:general_cp}). By means of quasi-symbolic calculations we also show  that local clustering depends on the expected node degree $k$ similarly for both lower and upper bounds of the effective connection probability, and that average clustering does not depend on average degree $\bar k$.

Let the hyperbolic polar coordinates of the point triplet
$u,v,w$ be $(r_u,\phi_u),(r_v,\phi_v),(r_w,\phi_w)$ and $\phi=\phi_u-\phi_v$,
$\psi=\phi_u-\phi_w$. The local clustering coefficient $cl(r_u)$ for a given
node $u$ is calculated as the ratio of the expected number of link pairs with
common edge $u$ and the expected number of link triangles with edge $u$. For
calculating these expected numbers, the joint probabilities of the existence of
$(u,v)$ and $(u,w)$ link pair and the existence of the $(u,v,w)$ link triangle
are substituted by $p(d_{uv}) p(d_{uw})$ and $p(d_{uv}) p(d_{uw}) p(d_{vw})$,
respectively. This requires link independence assumption, which is not true,
however, correlations are expectedly diminished due to averaging processes (like in
mean field calculations \cite{meanfieldcluster}). In this way, the local clustering coefficient is
formulated as
%\footnote{Note, that in the formulae of expected link pairs and
%triangles $\delta^2$ factor is present, which disappears in their ratio.}

\begin{equation}
cl(r_u)=\frac{\delta^2 \int_{r_w=0}^R \int_{r_v=0}^R \int_{\psi=0}^{2 \pi}
\int_{\phi=0}^{2 \pi} p(d_{uv}) p(d_{uw}) p(d_{vw}) {\rm d} \phi {\rm d} \psi
\sinh(r_v) \sinh(r_w) {\rm d} r_v {\rm d} r_w }{\delta^2 \int_{r_w=0}^R
\int_{r_v=0}^R \int_{\psi=0}^{2 \pi} \int_{\phi=0}^{2 \pi} p(d_{uv}) p(d_{uw})
{\rm d} \phi {\rm d} \psi \sinh(r_v) \sinh(r_w) {\rm d} r_v {\rm d} r_w } \ .
\end{equation}
For estimating these integrals in the numerator and the denominator the
following functions are defined:
\begin{eqnarray} \label{equ:Nuxyz}
& & \int_{\psi=0}^{2 \pi} \int_{\phi=0}^{2 \pi} p(d_{uv}) p(d_{uw}) p(d_{vw}) {\rm
d} \phi {\rm d} \psi \approx \\
&\approx& \int_{\psi=0}^{2 \pi} \int_{\phi=0}^{2 \pi}
\exp\left(-x \sin \frac{\phi}{2}-y \sin \frac{\psi}{2}
 -z \sin \frac{|\psi-\phi|}{2} \right) {\rm d} \phi {\rm d} \psi =: {\rm Nu}(x,y,z) \nonumber
\end{eqnarray}
and
\begin{equation}
\int_{\psi=0}^{2 \pi} \int_{\phi=0}^{2 \pi} p(d_{uv}) p(d_{uw})
 {\rm d} \phi {\rm d} \psi \approx \int_{\psi=0}^{2 \pi} \int_{\phi=0}^{2 \pi}
\exp\left(-x \sin \frac{\phi}{2}-y \sin \frac{\psi}{2} \right) {\rm d} \phi
{\rm d} \psi =: {\rm De}(x,y)
\end{equation}
where the general connection probability formula  (\ref{equ:general_cp}), the
approximation $e^{\frac{d_{uv}}{2}} \approx
e^{\frac{r_u+r_v}{2}} \sqrt{\frac{1-\cos \phi}{2}}$ are applied and
\begin{equation} \label{equ:xyzdef}
x = \frac{8}{\bar k} \delta e^{\frac{r_u+r_v}{2}} \ , \ y = \frac{8}{\bar k} \delta e^{\frac{r_u+r_w}{2}} \ ,
\ z = \frac{8}{\bar k} \delta e^{\frac{r_v+r_w}{2}} \ .
\end{equation}

Now we apply asymptotic expansions of Nu$(x,y,z)$
and De$(x,y)$ in order to approximate them. (Asymptotic expansion here means
that $x,y,z$ are large parameters and we are interested in the asymptotic
behaviour of these integrals as $\{x,y,z\} \rightarrow \infty$). Note that De$(x,y)$ is simply the
product of two integrals which reads as
\begin{eqnarray}
{\rm De}(x,y):= \int_{\psi=0}^{2 \pi} \exp\left(-y \sin \frac{\psi}{2} \right)
 {\rm d} \psi\int_{\phi=0}^{2 \pi} \exp\left(-x \sin \frac{\phi}{2} \right) {\rm d}
 \phi \nonumber \\ = 4 \pi^2 ( {\rm I}(0,x)-{\rm S}(0,x)) ( {\rm I}(0,y)-{\rm S}(0,y))
 \approx \frac{16}{x y}
\end{eqnarray}
due to that I$(0,x)$-S$(0,x)$ $\approx \frac{2}{\pi} x^{-1}$ based on its
asymptotic expansion \cite{abramovitz} .
%(see also equation (\ref{eqn:besselstruvel})).

For approximating ${\rm Nu}(x,y,z)$ we use Laplace's \cite{laplacemethod}
method to generate first orders of the asymptotic expansion with respect to
$x,y$ and $z$. For this we take the first order Taylor series expansion of the
sinus functions around $0$ and $2 \pi$ where the integral is dominant for
larger $x,y,z$. Performing the double integral (\ref{equ:Nuxyz}) with these
series and erasing the exponentially small terms, we get the following four terms with respect to that $x$ is in the neighbourhood of $0$ or $2 \pi$ and $y$ is in the neighbourhood of $0$ or $2 \pi$ :

%\begin{eqnarray}
%\!\!\!\!\!\!\!\!\!\!\!\!\!\!\!\!\!\!\!\!\!\!\!\!\!\!\!\!\!\!\!\!\!\!\!\!\!\!\!\!\!\!\!\!\!\!\!\frac{-\left(4
%e^{-2 \pi (x+y+z)} \left(-e^{\pi (2 x+y+z)} (x+y) (y-z) (x+z)+ e^{\pi  (x+2
%y+z)} (x+y) (-x+z) (y+z)+e^{\pi (x+y+2 z)} (x+y-2 z) (x+z) (y+z)-e^{2 \pi
%(x+y+z)} (y-z) (-x+z) (x+y+2 z)\right)\right)}{((x+y) (x-z) (x+z) (-y+z)
%(y+z))}
%\end{eqnarray}
%\begin{eqnarray}
%\frac{-\left(4
%e^{-2 \pi (x+y+z)} \left(-e^{\pi (2 x+y+z)} (x+y) (y-z) (x+z)+ e^{\pi  (x+2
%y+z)} (x+y) (-x+z) (y+z)\right)\right)}{((x+y) (x-z) (x+z) (-y+z) (y+z))} + \nonumber \\
%+\frac{e^{\pi (x+y+2 z)} (x+y-2 z) (x+z) (y+z)-e^{2 \pi (x+y+z)} (y-z)
%(-x+z) (x+y+2 z)}{((x+y) (x-z) (x+z) (-y+z) (y+z))}
%\end{eqnarray}
%and erasing the exponentially smaller terms we obtain the following simple
%formula:
\begin{equation}
{\rm Nu}(x,y,z) \approx 2 \frac{4(x+y+2 z)}{(x+y) (x+z) (y+z)} + 2 \frac{4}{(x+z)(y+z)} = \frac{16 (x+y+z)}{(x+y)(x+z)(y+z)} \ .
\end{equation}
Now the clustering coefficient can be written as
\begin{eqnarray}
cl(r_u) \approx \frac{\frac{\delta^2}{2} \int_{r_w=0}^R \int_{r_v=0}^R {\rm
Nu}(x,y,z) \sinh(r_v) \sinh(r_w) {\rm d} r_v {\rm d} r_w }{\frac{\delta^2}{2}
\int_{r_w=0}^R \int_{r_v=0}^R {\rm De}(x,y,z) \sinh(r_v) \sinh(r_w) {\rm d} r_v
{\rm d} r_w} \approx \nonumber \\ \approx \frac{\int_{r_w=0}^R \int_{r_v=0}^R \frac{16(x+y+z)}{(x+y)
(x+z) (y+z)} \sinh(r_v) \sinh(r_w) {\rm d} r_v {\rm d} r_w}{\int_{r_w=0}^R
\int_{r_v=0}^R \frac{16}{x y} \sinh(r_v) \sinh(r_w) {\rm d} r_v {\rm d} r_w} \label{eqn:clru}
\end{eqnarray}
%\footnote{Note that ${\rm Nu}(x,y)$ and ${\rm De}(x,y)$ are approximately upper
%bounds on the double integrals by angles in the numerator and denominator,
%however, the ratio of the double integration by radial coordinates of these
%functions appears in the clustering coefficient approximations, which is not
%necessarily an upper bound.}
Based on this it can be seen that $cl(r_u)$ does
NOT depend on the density parameter $\delta$, and depends on the average degree $\bar k$  only through $r_u(k,{\bar k})$ (see equation (\ref{equ:general_outdegree}) ) because all the $x,y,z$ terms
contain a $\frac{8}{\bar k} \delta$ factor. In this way both integrals in the numerator and
denominator posses a $\frac{1}{\delta^2}$ factor. (Note, that both the
numerator and denominator are independent from $\delta$).

In what follows we explore how the local clustering coefficient of a node is
depending on the expected degree $k$. This is possible to perform through the
inverse function of ${\bar k}(r_u)$ (based on (\ref{equ:general_outdegree})) which is
$
r_u(k) =  R-2 \ln \left({2 k}/{\bar{k}}\right) \ .
$
%$r_u(k)= 2 \ln
%\left(\frac{1}{k}\right)+2 \ln (2)+R$, see also (\ref{eqn:kruapproximation}).
First the denominator is calculated which is possible in a parametric way.
\begin{equation} \label{eqn:k2}
\int_{r_w=0}^R \int_{r_v=0}^R \frac{16}{x y} \sinh(r_v) \sinh(r_w) {\rm d} r_v
{\rm d} r_w = \frac{1}{9} e^{-4 R} (1 - 4 e^{3 R/2} + 3 e^{2 R})^2 k^2 \approx
k^2 
\end{equation}
with the substitutions $x,y$ in (\ref{equ:xyzdef}) and $r_u(k)$ above. (The term $\frac{16}{x y}$ does not depend on $\bar k$ due to the $x,y$ and $r_u(k)$ substitution). Note
that this is a good cross-validation of this formula, because the expected
number of link pairs of a node with given expected degree $k$ is approximately
$k (k-1)/2 \approx k^2/2$. This is because if the node degree $\kappa$ has Poisson
distribution with parameter $k$ then the expected number of link pairs at this
node is $E\left[\frac{\kappa (\kappa-1)}{2}\right] = \sum_{l=0}^\infty
\frac{l(l-1)}{2} \frac{k^l}{l!} e^{-l}$, which is exactly $\frac{k^2}{2}$.
Based on the equations (\ref{eqn:clru}), (\ref{eqn:k2}) and substituting
$x,y,z$ into the formula of the integrand one can obtain
%\begin{equation}
%cl(k) \approx \int_{r_w=0}^R \int_{r_v=0}^R \frac{e^{\frac{1}{2}
%\left(-R+r_v+r_w\right)} \left(e^{\frac{1}{2}
%\left(R+r_v\right)}+e^{\frac{1}{2} \left(R+r_w\right)}+e^{\frac{1}{2}
%\left(r_v+r_w\right)} k\right)}{4
%\left(e^{\frac{r_v}{2}}+e^{\frac{r_w}{2}}\right) \left(2
%e^{R/2}+e^{\frac{r_v}{2}} k\right) \left(2 e^{R/2}+e^{\frac{r_w}{2}}
%k\right)}{\rm d} r_v {\rm d} r_w
%\end{equation}
\begin{equation} \label{equ:clkintegral}
%cl(k) \approx \int_{r_w=0}^R \int_{r_v=0}^R \frac{\bar{k} e^{\frac{1}{2} \left(r_v+r_w-R\right)}
%\left(\bar{k} e^{\frac{1}{2} \left(r_v+R\right)}+\bar{k} e^{\frac{1}{2}
%\left(r_w+R\right)}+4 k e^{\frac{1}{2} \left(r_v+r_w\right)}\right)}{16
%\left(e^{\frac{r_v}{2}}+e^{\frac{r_w}{2}}\right) \left(e^{R/2} \bar{k}+2 k e^{\frac{r_v}{2}}\right)
%\left(e^{R/2} \bar{k}+2 k e^{\frac{r_w}{2}}\right)} {\rm d} r_v {\rm d} r_w
cl(k,\bar k, R) \approx \int_{r_w=0}^R \int_{r_v=0}^R \frac{\bar{k} e^{\frac{1}{2} \left(r_v+r_w-R\right)} \left(\bar{k} e^{\frac{1}{2} \left(r_v+R\right)}+\bar{k} e^{\frac{1}{2} \left(r_w+R\right)}+2 k e^{\frac{1}{2} \left(r_v+r_w\right)}\right)}{4 \left(e^{\frac{r_v}{2}}+e^{\frac{r_w}{2}}\right) \left(e^{R/2} \bar{k}+2 k e^{\frac{r_v}{2}}\right) \left(e^{R/2} \bar{k}+2 k e^{\frac{r_w}{2}}\right)} {\rm d} r_v {\rm d} r_w
\ .
\end{equation}
This double integral on the right hand side can be assessed symbolically by substitution,
but even a simplified result is still quite spacious (see the next note). Nevertheless, the detailed analysis of this function reveals that it is approximately independent of $R$, and as $k$ is increasing, the local clustering coefficient tends to
\begin{equation}
cl(k,\bar k) \approx \ln (2) {\bar k} k^{-1} \ .
\end{equation}

For simplicity and for catching the behaviour of $cl(k,\bar k)$ even for smaller $k$ values, the following intuitive form of approximation is calculated by numerical matching. The intuition is based on the observation that the integrand itself is in the form of a fraction of a first order and a second order polynomial of $k$.
%for any fixed $\bar k$ a good numerical matching can be calculated in the form
\begin{equation}\int_{r_w=0}^R \int_{r_v=0}^R \frac{\bar{k} e^{\frac{1}{2} \left(r_v+r_w-R\right)} \left(\bar{k} e^{\frac{1}{2} \left(r_v+R\right)}+\bar{k} e^{\frac{1}{2} \left(r_w+R\right)}+2 k e^{\frac{1}{2} \left(r_v+r_w\right)}\right)}{4 \left(e^{\frac{r_v}{2}}+e^{\frac{r_w}{2}}\right) \left(e^{R/2} \bar{k}+2 k e^{\frac{r_v}{2}}\right) \left(e^{R/2} \bar{k}+2 k e^{\frac{r_w}{2}}\right)} {\rm d} r_v {\rm d} r_w \ \approx \frac{1+ a k}{b+c k+ d k^2}
\end{equation}
%% a fenti ket keplet mar a javitott kepletek az uj 16 (x+y+z)/((x+y)(x+z)(y+z))) keplettel szamolva, ld szinten a math filet clustering_coefficient_recalculated
where the coefficient $a,b,c,d$ are approximately independent of $R$
and is depending only on $\bar k$.  The coefficient is summarised in
the Table~\ref{table:abcd} for three cases: for the lower
bound of the average degree $1$, for the upper bound $4$, and $\bar{k}
= 2.27$ which latter average degree comes from the numerical
simulation of the network formation game.

\begin{table}[h]
  \centering
\begin{tabular}{|r||r|r|r|r|}
  \hline
  $\bar k$ & a & b & c & d \\
  \hline \hline
  1     & 0.598 & 1.008 & 2.168 & 0.869 \\ \hline
  2.27  & 0.331 & 1.002 & 1.019 & 0.209 \\ \hline
   4    & 0.220 & 1.002 & 0.618 & 0.080 \\
  \hline
\end{tabular}
\caption{The clustering coefficient as a function of the average degree.}
\label{table:abcd}
\end{table}

Note that, for larger $k$'s
\begin{equation}\label{eqn:clustering_vs_degree}
\frac{1+ a k}{b+c k+ d k^2} \approx \frac{a}{d} k^{-1} \ , \ \frac{1}{2} \bar k \leq k \leq \frac{1}{2} \bar k e^{\frac{R}{2}}  \ ,
\end{equation}
and $\frac{a}{d}$ is very close to $\ln(2) {\bar k}$ for all the three cases, as expected.

\begin{figure}[h!]
\psfrag{Delta}{$\delta$}
\psfrag{Mod}{NNG sim.}
\psfrag{Tha}{Theory (\ref{eqn:delta_clustering})}
\psfrag{Thb}{Theory (\ref{eqn:clustering_vs_degree})}
\begin{center}
    \includegraphics[width=.95\textwidth,angle=0]{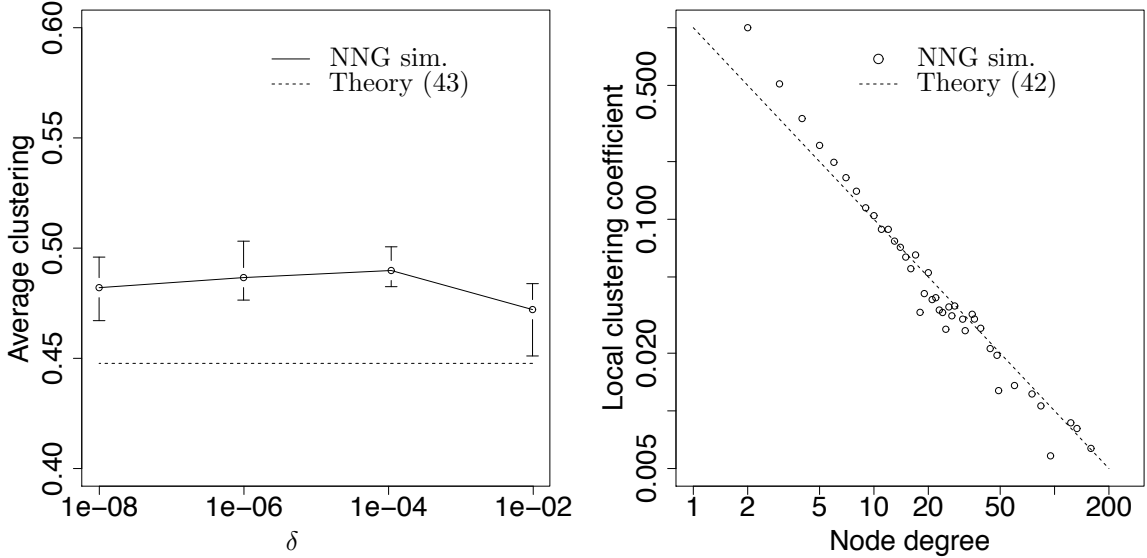}
     \caption{Average clustering as a function of $\delta$, and local clustering as a function of node degree.}
    \label{fig:clustering_vs_delta}
\end{center}
\end{figure}

It is now possible to compute average clustering based on the
approximation above as
\begin{equation} \label{eqn:delta_clustering}
cl = \int_{k=\frac{1}{2} \bar k}^{\frac{1}{2} \bar k e^{\frac{R}{2}}} cl(k,\bar k ) \frac{\partial}{\partial k} (1-\bar F(k)) {\rm d} k
\approx  \int_{k=\frac{1}{2} \bar k}^{\frac{1}{2} \bar k e^{\frac{R}{2}}} \frac{1+ a k}{b+c k+ d k^2} \frac{\bar k^2}{ 2 k^3} {\rm d} k \ .
\end{equation}
Evaluating this integral for the average degree lower bound
$\bar{k}=1$, upper bound $\bar{k}=4$, and the average degree in
simulations $\bar{k} = 2.27$, we obtain, using Table~\ref{table:abcd},
$cl=0.447075, 0.447615, 0.447146$, respectively. We have also
performed more extensive numerical experiments showing that average
clustering does not significantly depend on the average degree for
$\delta\in[10^{-8},10^{-2}]$ and $R\in[10,20]$.  Its dependence on $R$
is also negligible, which is not surprising since $R$ appears only on
the upper limit of the integral, and this upper limit negligibly
affects the result since the integrands decrease as $\sim k^{-5}$. All
these analytic and numeric results are in a good agreement with
simulations, see Figure~\ref{fig:clustering_vs_delta}.

\begin{centering}
  \section{Appendix 8 - Evaluating the integral (\ref{equ:clkintegral})}
\end{centering}
The integral for computing the local clustering coefficient presented (\ref{equ:clkintegral}) can be evaluated by the following substitution
\begin{equation}
\xi = \Exp\left( \frac{r_v}{2} \right) \  , \ \zeta = \Exp \left( \frac{r_w}{2} \right) \  , \
{\rm d} \xi = \Exp\left( \frac{r_v}{2} \right) \frac{1}{2} {\rm d} r_v \ , \
{\rm d} \zeta =  \Exp\left( \frac{r_w}{2} \right) \frac{1}{2} {\rm d} r_w \ ,
\end{equation}
which is in the form
\begin{equation}
%cl(k) \approx \int_{r_w=0}^R \int_{r_v=0}^R \frac{\bar{k} e^{\frac{1}{2} \left(r_v+r_w-R\right)}
%\left(\bar{k} e^{\frac{1}{2} \left(r_v+R\right)}+\bar{k} e^{\frac{1}{2}
%\left(r_w+R\right)}+4 k e^{\frac{1}{2} \left(r_v+r_w\right)}\right)}{16
%\left(e^{\frac{r_v}{2}}+e^{\frac{r_w}{2}}\right) \left(e^{R/2} \bar{k}+2 k e^{\frac{r_v}{2}}\right)
%\left(e^{R/2} \bar{k}+2 k e^{\frac{r_w}{2}}\right)} {\rm d} r_v {\rm d} r_w
cl(k,\bar k, R) \approx \int_{1}^{e^{\frac{R}{2}}} \int_{1}^{e^{\frac{R}{2}}}
\frac{e^{-\frac{R}{2}} \bar{k} \left(e^{\frac{R}{2}} \bar{k} (\zeta +\xi )+2 \zeta  k \xi \right)}{(\zeta +\xi ) \left(e^{\frac{R}{2}} \bar{k}+2 \zeta  k\right) \left(e^{\frac{R}{2}} \bar{k}+2 k \xi \right)} {\rm d} \xi {\rm d} \zeta \ .
\end{equation}
A simplified version of the result of the integral (\ref{equ:clkintegral}) is
\begin{dmath}
\frac{1}{8 k^2} \left( e^{-\frac{R}{2}} \left(\bar{k} e^{R/2} \left(\bar{k} \left(\text{Li}_2\left(\frac{2 \left(1+e^{-\frac{R}{2}}\right) k}{2 k-\bar{k}}\right)-\text{Li}_2\left(\frac{2 \left(1+e^{-\frac{R}{2}}\right) k}{2 k+\bar{k}}\right)+
\\
\text{Li}_2\left(-\frac{2 \left(1+e^{R/2}\right) k}{e^{R/2} \bar{k}-2 k}\right)+\text{Li}_2\left(\frac{4 k}{2 k+e^{R/2} \bar{k}}\right)-\text{Li}_2\left(\frac{2 \left(1+e^{R/2}\right) k}{2 k+e^{R/2} \bar{k}}\right)-\text{Li}_2\left(\frac{4 k}{2 k-e^{R/2} \bar{k}}\right)-
\\
\text{Li}_2\left(-\frac{4 k}{\bar{k}-2 k}\right)+\text{Li}_2\left(\frac{4 k}{2 k+\bar{k}}\right)\right)+\bar{k} \left(\log \left(e^{R/2}+1\right) \left(\ln \left(-\frac{2 k e^{-\frac{R}{2}}+\bar{k}}{2 k-\bar{k}}\right)-\ln \left(\frac{\bar{k}-2 k e^{-\frac{R}{2}}}{2 k+\bar{k}}\right)+
\\
\ln \left(\frac{e^{R/2} (2 k+\bar{k})}{\bar{k} e^{R/2}-2 k}\right)-\ln \left(\frac{e^{R/2} (\bar{k}-2 k)}{2 k+\bar{k} e^{R/2}}\right)\right)+\ln \left(2 e^{R/2}\right) \left(\ln \left(1-\frac{4 k}{2 k+\bar{k}}\right)-\ln \left(\frac{4 k}{\bar{k}-2 k}+1\right)\right)+
\\
\ln (2) \left(\ln \left(1-\frac{4 k}{2 k+\bar{k} e^{R/2}}\right)-\ln \left(\frac{4 k}{\bar{k} e^{R/2}-2 k}+1\right)\right)+2 \left(\ln \left(e^{R/2} (2 k+\bar{k})\right)-
\\
\ln \left(2 k+\bar{k} e^{R/2}\right)\right) \left(\tanh ^{-1}\left(\frac{2 k}{\bar{k}}\right)-\tanh ^{-1}\left(\frac{2 k e^{-\frac{R}{2}}}{\bar{k}}\right)\right)\right)+8 k \left(\ln \left(e^{R/2}\right)-\ln \left(e^{R/2}+1\right)+\ln (2)\right)\right)+
\\
4 k \bar{k} \left(\ln (4)-2 \ln \left(e^{R/2}+1\right)\right)\right)\right) ,
\end{dmath}
%\begin{equation}
%\begin{split}
%
%\end{split}
%\end{equation}
where the function $\text{Li}_2(z) = \sum _{k=1}^{\infty } \frac{z^k}{k^2}$ is the di-logarithm special function.
We observe that factors $\Exp(-R/2)$ and $\Exp(R/2)$ appear in several terms. If $R$ is sufficiently large, e.g., ranging between realistic values of $10$ and $20$, then we can neglect the exponentially smaller terms, keeping only the exponentially large dominating terms.
For example,
\begin{equation}
\frac{\bar{k}-2 k e^{-\frac{R}{2}}}{2 k+\bar{k}} \approx \frac{\bar k }{2k+\bar k} \ \ \ {\rm and} \ \ \
\frac{e^{R/2} (2 k+\bar{k})}{\bar{k} e^{R/2}-2 k} \approx \frac{2k+\bar k}{\bar k} \ .
\end{equation}
Using this procedure, after some simplifications, we finally obtain an $R-$free expression for clustering:
\begin{eqnarray}
cl(k,\bar k) \approx \frac{1}{8 k^2}\bar{k}\left(8 k \ln (2)+\bar{k} \left(\ln \left(\frac{\bar{k}+2 k}{\bar{k}}\right) \ln \left(\frac{\bar{k}+2 k}{\bar{k}-2 k}\right)+\ln (2) \ln \left(\frac{\left(\bar{k}-2 k\right)^2}{\left(\bar{k}+2 k\right)^2}\right)\right)+\right. \nonumber \\
\bar{k}\left(\text{Li}_2\left(\frac{2 k}{2 k-\bar{k}}\right)+\text{Li}_2\left(-\frac{2 k}{\bar{k}}\right)-\text{Li}_2\left(\frac{2 k}{\bar{k}}\right)-\right. \nonumber \\
\left.\left.\text{Li}_2\left(-\frac{4 k}{\bar{k}-2 k}\right)-\text{Li}_2\left(\frac{2 k}{2 k+\bar{k}}\right)+\text{Li}_2\left(\frac{4 k}{2 k+\bar{k}}\right)\right)\right)
\end{eqnarray}
We can now see that $cl(k,\bar k) \rightarrow \ln(2) \bar k \ k^{-1}$ as $k$ increases, because the logarithmic terms become zero, while the dilogarithmic terms eliminate each other.
The analysis of this function at $k=0$ also shows that $cl(0,\bar k)=1$, from which it follows that $b=1$ in the polynomial matching the numerical calculations, cf.\ Table~\ref{table:abcd}.

%\begin{centering}
%  \section{Appendix Note 11 - Addition of edges according to
%    betweenness centrality}
%\end{centering}
%
%\begin{fiure}[H]
%\begin{center}
%    \includegraphics[width=.45\textwidth,angle=0]{ncorr_bet.ps}
%     \caption{Addition of edges based on betweenness
%       values versus from NNG.}
%    \label{fig:add_betw}
%\end{center}
%\end{figure}

\begin{centering}
\section{Appendix 9 - Expected out-degree distribution in a frame topology with quasi-uniform node density}
\end{centering}

The radial coordinate density in case of quasi-uniform node density is
\begin{equation}
\rho(r,\alpha):=\frac{\alpha  \sinh (\alpha  r)}{\cosh (\alpha  R)-1} \approx \alpha e^{\alpha (r-R)}
\end{equation}
while the angle density remains uniform ($\frac{1}{2 \pi}$) over the range $[1,2 \pi]$.
Given a point pair $(u,v)$, first we determine the probability $p(r_u,\alpha)$ that the $u \rightarrow v$ link exists, then based on this the average out degree $k(r_u,\alpha)$ of $u$ is calculated, and finally $\bar F(k,\alpha)$ is also given.

Probability $p(r_u,\alpha)$ is equal to the probability that none of the remaining
$N-2$ points fall in the intersection of the $v$-centred $d_{uv}$
circle and the $R$-disk. Let us denote by $p_1$ the probability that a point whose coordinates generated by randomly according to the densities above falls inside the intersection. Using $p_1$ the probability $p(r_u,\alpha)$ can be calculated and approximated as
\begin{equation}
p(r_u,\alpha) = (1-p_1)^{N-2} \approx e^{-N p_1}
\end{equation}
%% to do labjegyzetbe beirni, hogy ez akkor pontos, ha n p_1< 10, de ez pont jo nekunk, mert az integralsaoknal a kicsi N p_1 tartomanyok lesznek a dominansak...
The calculation of $p_1$ can be performed by using the node density function in the following way \cite{dima_infocom_2010}
\begin{equation}
p_1= \int_0^{\max(0,d-r_v)}  \rho(r,\alpha) {\rm d} r + \frac{1}{2 \pi}\int_{|d-r_v|}^{\min(R,d+r_v)} \rho(r,\alpha) 2 \theta(r) {\rm d} r
\end{equation}
where
\begin{equation}
\theta(r)=\arccos \frac{\cosh r_v \cosh r-\cosh d}{\sinh r_v \sinh r } \ .
\end{equation}
In \cite{dima_infocom_2010} a useful approximation is presented for quite similar integrals, based on which one can write
\begin{equation}
p_1 \approx \frac{4 e^{\frac{1}{2} (d-R-r_v)} \alpha }{\pi  (-1+2 \alpha )}
\end{equation}
for $0.5 < \alpha \leq 1$ .
%% to do esetleg labjegyzet, hogy alpha > .7 eseten a hiba kisebb mint 6%.

Now the expected out-degree of $u$ can be written as
\begin{equation}
k_{\mathrm{out}}(r_u,\alpha) \approx \frac{N}{2 \pi} \int_0^R  \int_0^{2 \pi} e^{-N p_1}  {\rm d} \phi \rho(r_v) {\rm d} r_v \ .
\end{equation}
Using the approximation of $p_1$ and $\cosh(d/2) \approx e^{\frac{r_u+r_v}{2}} \sin \frac{\phi}{2}$ %(see equation (\ref{equ:ed2approx}))
one can formulate
\begin{equation}
\int_0^{2 \pi} e^{-N p_1}  {\rm d} \phi \approx \int_0^{2 \pi} e^{-x \sin \frac{\phi}{2} }  {\rm d} \phi \approx 2 \pi ({\rm I} (0,x)-{\rm S}(0,x)) \approx \frac{4}{x}
\end{equation}
where
\begin{equation}
x = 4 \frac{N}{\pi} \frac{\alpha}{2 \alpha -1} e^{\frac{r_u-R}{2}} \ .
\end{equation}
Note, that $x$ does not depend on $r_v$, therefore the second integration by $r_v$ results
\begin{equation}
k_{\mathrm{out}}(r_u,\alpha) \approx \frac{N}{2 \pi} \frac{4}{x} \int_0^R \rho(r_v,\alpha) {\rm d} r_v =  \frac{2 \alpha-1}{2 \alpha} e^{\frac{R}{2}} e^{-\frac{r_u}{2}} \ .
\end{equation}
Note, that for $\alpha=1$ we get back the result for the uniform density case, (\ref{equ:general_outdegree}).

\begin{figure}[h!]
\psfrag{Delta}{$\delta$}
\psfrag{Mod}{NNG}
\psfrag{Theo}{Theory ($k^{-1}$)}
\psfrag{no}{$\alpha=0.5$}
\psfrag{nho}{$\alpha=0.75$}
\psfrag{ne}{$\alpha=1$}
\psfrag{k}{$k$}
\psfrag{F(k)}{$\bar{F}(k)$}
\psfrag{slok}{Slope $-2$}
\psfrag{sloe}{Slope $-1.1$}
\begin{center}
    \includegraphics[width=.95\textwidth,angle=0]{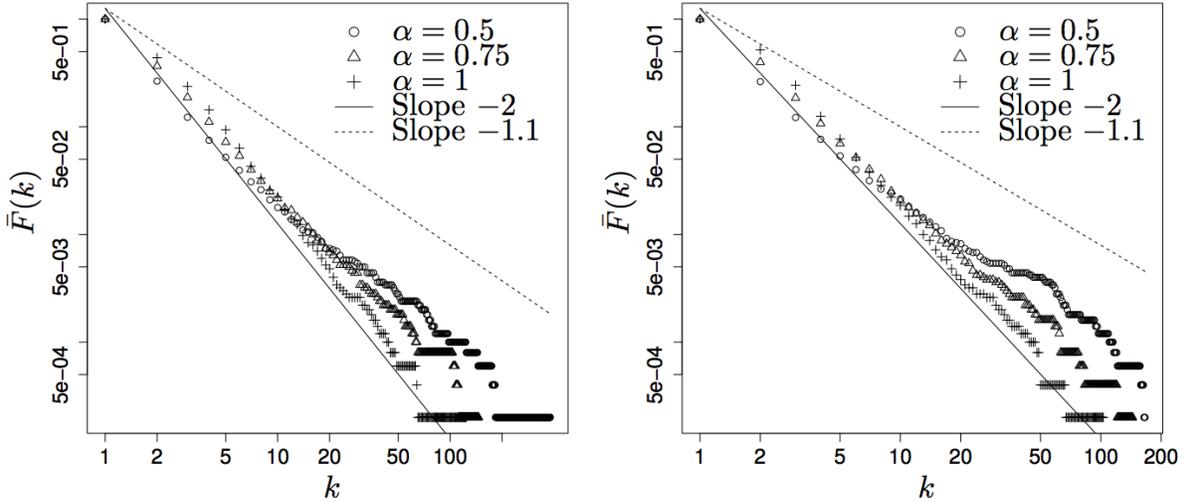}
     \caption{The in and out degree distributions of the NNG for various settings of the $\alpha$ parameter.}
    \label{fig:dd_vs_alpha}
\end{center}
\end{figure}

Now the (approximation of the) complement cumulative distribution function $\bar F (k)$ can be derived as,
\begin{equation}
\bar F (k) = \int_0^{r_u(k)} \rho(r,\alpha) {\rm d} r \approx e^{\alpha (r_u(k)-R)}  = \left(\frac{1-\frac{1}{2 \alpha }}{k}\right)^{2 \alpha }
\end{equation}
where $r_u(k)$ is the inverse function of $k_{\mathrm{out}}(r_u)$ . The
simulation results displayed in Figure~\ref{fig:dd_vs_alpha} readily
confirm this finding.

\begin{centering}
  \section{Appendix 10 - Statistical Significance}
\end{centering}

In this note we provide probability estimates which represent the statistical significance of that the NNG equilibrium network links' containment by the real networks is very unlikely to occur by random chance, but rather is likely to be attributable to the specific characteristics of our embedding and NNG processes.

The NNG equilibrium network (graph) is a transformation of the real network under investigation by an embedding and a gaming (NNG) process. Although this transformation is completely deterministic, the statistical significance test can be performed in the following two ways: In the first approach the NNG equilibrium network is substituted by a completely random network with the same average degree $\bar{k}_{\mathrm{NNG}}$, that is $\frac{N}{2} \bar{k}_{\mathrm{NNG}}$ links are randomly chosen from the possible $\frac{N(N-1)}{2}$ number of links. The probability that $p$ fraction of these links  (e.g. $p = 0.83$) are contained by the real network (having $\frac{N}{2} \bar{k}$ links) can be calculated as
\begin{equation}
\frac{\binom{N (N-1)/2-{N/2 \bar{k}}}{ (1-{p}) N/2 \bar k_{\mathrm{NNG}}} \binom{N/2 \bar k}{p N/2 \bar k_{\mathrm{NNG}}}}{\binom{N (N-1)/2}{N/2 \bar k_{\mathrm{NNG}}}}
\end{equation}
 which is in the order of $O(e^{-N})$. Because this probability is extremely small for reasonable $N$, our result is very unlikely to occur also along with fully random networks with fixing only the number of edges. For example, taking the values on the Internet AS-level topology embedding ($N= 4919, \frac{N}{2} \bar k = 28361, \frac{N}{2} \bar k_{\mathrm{NNG}} = 5490, p=0.83$) the probability above is $5.62\times10^{-11068}$.

 A more refined randomization of the NNG equilibrium network is to substitute only the embedding process by fully random generation of H2 coordinates (with such coordinate distribution similar to the one resulted by the embedding process) and then apply the gaming process (as if the embedding was wrong and had no concern to the original real network).  In this way, the resulted random NNG network preserves not only the average degree, but the degree distribution and also the clustering coefficient of the original NNG equilibrium network.
 %% ide kene még kis okoskodas
 Let $X$ be a random variable denoting the number of links from the randomized NNG equilibrium network contained by the original real network. Inevitably, $X$ is a non-negative random variable bounded also from above by $P:=\frac{N}{2} \bar k_{\mathrm{NNG}}$.
 Although the exact distribution of $X$ cannot be calculated due to the dependent  link establishment of the gaming process, the expected value of $X$ (which is insensitive to link dependence) is
 \begin{equation}
 E(X) = \frac{N}{2} \bar k \frac{\frac{N}{2} \bar k_{\mathrm{NNG}}}{\frac{N(N-1)}{2}} \approx \frac{1}{2} \bar k_{\mathrm{NNG}} \, \bar k.
 \end{equation}
 Based on this average value, a conservative upper bound can also be
 given on the probability that the level of this link containment
 exceeds a certain threshold $0 < C < P$. Applying Hoeffding's
 inequality~\cite{hoeffding1963probability} we can state that
 \begin{equation}
 P(X > C) \leq \left( \frac{E(X)}{C} \right)^{\frac{C}{P}} \left(\frac{P-E(X)}{P-C}\right)^{1-\frac{C}{P}}
 \end{equation}
 This upper bound is far below 0.05 for several reasonable $\bar k$ and $N$. For example, the probability that more than 83 percent of the randomized NNG equilibrium network links ($C=$4556 of the total 5490 edges) coincide Internet real edges (among the total 28 361) is upper bounded by 0.00136044. The complement of the upper bound of the probability above (1-upper bound) can also be considered as a weight of our statement (in the example above 0.99864).\\

 %These numbers confirm that the NNG equilibrium network links are to a great extent contained by the original networks) is very unlikely to occur by random chance, but rather is likely (by at least the weight) to be attributable to the specific characteristic of our embedding and NNG processes.\\

\begin{centering}
  \section{Appendix 11 - Euclidean plane}
\end{centering}

In this note we analyze the degree distribution in NNG equilibrium networks constructed on sets of points sprinkled uniformly at random over Euclidean disks.
We show that the expected degree of a node located in the disk centre is around $1$, while the expected degree of a node at the disk boundary is around $1/2$.
In view of this lack of variability of node degrees, the degree distribution in the Euclidean case cannot have any fat tails.

According to (\ref{equ:Tuvforkru}) the expected degree of a node $u$ is

\begin{equation}
\delta \int_0^R \int_{0}^{2 \pi} e^{-\delta T_{uv}} {\rm d} \phi r_v  {\rm d} r_v \ ,
\end{equation}
where $\delta=N/T_R=\frac{N}{R^2\pi}$. To give an upper bound we will give a lower bound for $T_{uv}$. If $u$ is the centre of the disk, then $T_{uv}$ is the area of the intersection of the disk and an circle around $v$ with radius $r_v$. If $r_v\le R/2$, then this intersection is the circle itself around $v$, else the intersection contains a circle with radius $R/2$, hence
\begin{multline}
k(0)\le \delta \int_0^{R/2} \int_{0}^{2 \pi} e^{-\delta r_v^2\pi} {\rm d} \phi r_v  {\rm d} r_v + \delta \int_{R/2}^R \int_{0}^{2 \pi} e^{-\delta (R/2)^2\pi} {\rm d} \phi r_v  {\rm d} r_v \\
\le 1-e^{-\frac14 \delta R^2\pi}+\frac34\delta R^2\pi e^{-\frac14 \delta R^2\pi} \le 1+3  \frac N4 e^{-N/4} \le 1+\frac3e.
\end{multline}
Moreover, if $N\ge6$ then $k(0)\le 1.05$

To give a lower bound to the expected degree we will count with the whole circle around $v$ instead of the intersection:

\begin{equation}
k(0)\ge \delta \int_0^{R} \int_{0}^{2 \pi} e^{-\delta r_v^2\pi} {\rm d} \phi r_v  {\rm d} r_v  = \delta 2\pi \int_0^{R}  e^{-\delta r_v^2\pi} r_v  {\rm d} r_v = 1-e^{-\delta R^2\pi} = 1-e^{-N}.
\end{equation}
If $N\ge6$, then $k(0)\ge0.99$.

Similarly, for the expected degree of a node $u$ at the disk boundary

\begin{equation}
k(R)\ge \delta \int_0^{R} \int_{0}^{2 \pi} e^{-\delta d^2\pi}{\rm d} \phi r_v  {\rm d} r_v,
\end{equation}
where $d$ is the distance between $u$ and $v$, and according to the cosines law, $d^2=R^2+r_v^2-2Rr_v\cos\phi_v$. The inner integration is
\begin{equation}
\int_{0}^{2 \pi} e^{-\delta\pi(R^2+r_v^2-2r_ur_v\cos\phi_v)} {\rm d} \phi = 2\pi {\rm I}(0,2\pi\delta r_v R)e^{-\delta\pi (R^2+r_v^2)},
\end{equation}
where ${\rm I}(0,x)$ is the BesselI function. Unfortunately the BesselI cannot be integrated, but we can use that ${\rm I}(0,x)\sim e^x/\sqrt{2\pi x}$. Hence
\begin{multline}
k(R)\ge \int_0^R \frac{2\pi\delta}{\sqrt{4\pi^2\delta R r_v}}e^{2\pi\delta R r_v-\delta\pi(R^2-r_v^2)}r_v  {\rm d} r_v= \int_0^R \frac{r_v N}{\sqrt{R^3 \pi}}e^{-\pi(R-r_v)^2N/R^2}  {\rm d} r_v \\
\ge \frac{2\sqrt N}{3\sqrt \pi}{\rm HypergeometricPFQ}\left(\left\lbrace\frac12,1\right\rbrace,\left\lbrace\frac54,\frac74\right\rbrace,-N\right) \xrightarrow[N\to\infty]{}\frac12
\end{multline}

\begin{figure}[h!]
\begin{center}
    \includegraphics[width=.95\textwidth,angle=0]{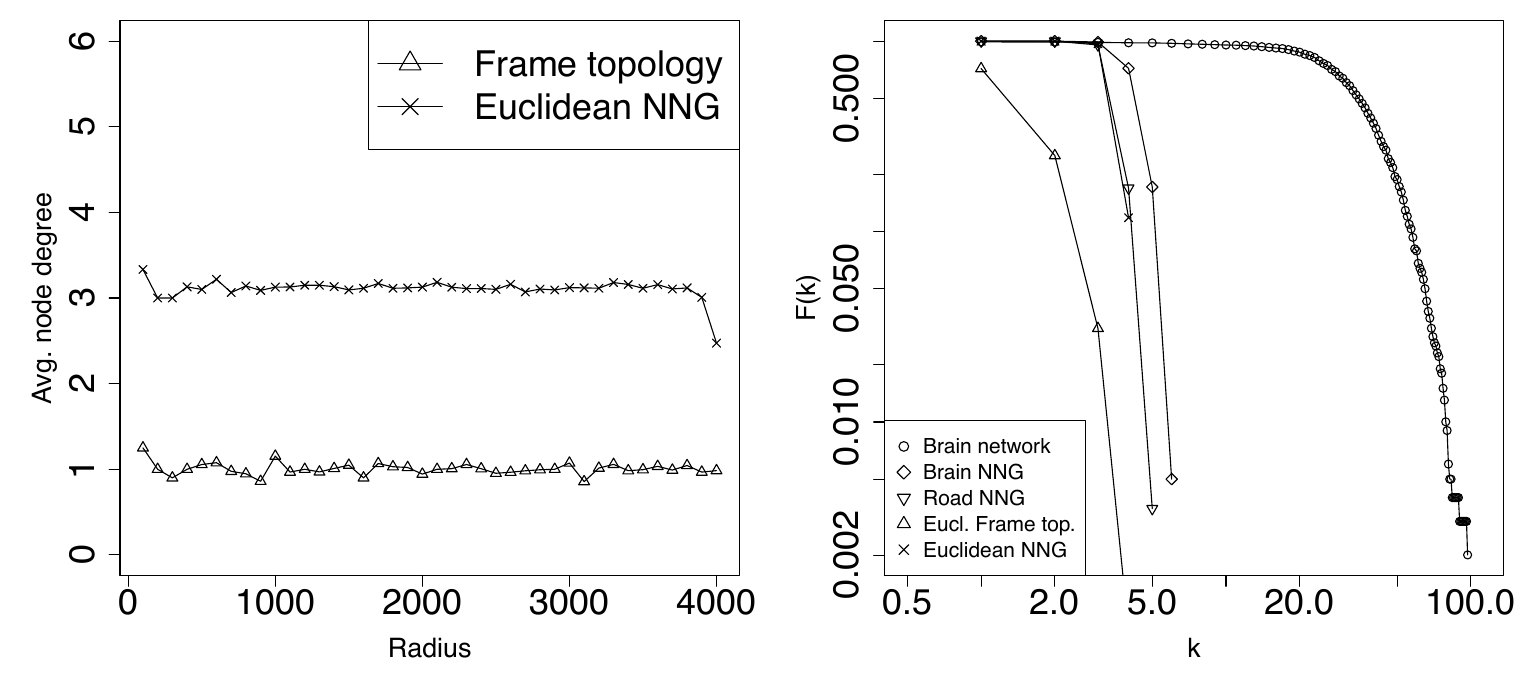}
    \caption{Average degree of nodes as a function of their radial coordinates on an Euclidean disk (left),
      and the cumulative distribution function of node degrees in the corresponding NNG equilibrium, its frame topology, the Hungarian road network, the brain network, and its NNG equilibrium (right).
      }
    \label{fig:euclidean}
\end{center}
\end{figure}

On the left panel of Figure~\ref{fig:euclidean} the simulation results
support the analytical findings that in the Euclidean case the
expected degree nodes as a function of their radial coordinates has
very low variability in the NNG equilibrium networks and their frame
topologies. As a consequence of this low variability the degree
distributions do not have any fat tails or power laws, and decay fast
with the node degree, the right panel of
Figure~\ref{fig:euclidean}. Clustering is still relatively strong
however: in the synthetic Euclidean NNG network it is $0.19$, in the road
NNG network it is $0.22$, while in the brain network and its NNG,
the clustering values are $0.46$ and $0.21$ respectively.

\begin{centering}
  \section{Appendix 12 - Heaviside step function approximation to the effective connection probability}
\end{centering}

The Heaviside step function with the step at
\begin{equation} \label{equ:Rprime}
R\rq{} = 2 \ln \frac{\bar k}{8\delta}
\end{equation}
is a good approximation to the effective connection probability in Eq.~(\ref{equ:general_cp}) for $\delta\in[10^{-6},10^{-3}]$ and $R=[12,18]$.
With this step-function approximation, node $u$ connects to $v$ iff $d_{uv} \leq R\rq{}$. Therefore the expected degree of $u$ is the expected number of points lying within the intersection of the $R-$disk and the $u$-centred disk of radius $R\rq{}$.

To see that this step function is indeed a good approximation to the effective connection probability in the NNG equilibrium, recall that the area of the two disks above can be approximated as
\begin{equation}
T_{R\rq{},R} = 4 e^{\frac{R\rq{}}{2}} e^{\frac{R-r_u}{2}}.
\end{equation}
From these one can obtain
\begin{equation}
k_{\mathrm{out}}(r_u) \approx N \frac{T_{R\rq{},R}}{T_{R-\mathrm{disk}}} = N \frac{4 e^{\frac{R\rq{}}{2}} e^{\frac{R-r_u}{2}}}{\pi e^{R}} \ .
\end{equation}
If $R\rq{}$ from (\ref{equ:Rprime}) is substituted into the formula above we get back the expected out-degree in (\ref{equ:general_outdegree}).
In particular, if $R\rq{}=R$ (as in \cite{krioukov2010hyperbolic}), then
\begin{equation}
k_{\mathrm{out}}(r_u) = \frac{4}{\pi} N e^{-\frac{r_u}{2}}
\end{equation}
and
\begin{equation}
\bar k = \frac{8}{\pi} N e^{-\frac{R}{2}},
\end{equation}
which coincides with Eqs.~(12,13) in \cite{krioukov2010hyperbolic}.\\

\begin{centering}
  \section{Appendix 13 - Nonnavigable network example}
\end{centering}

One cannot expect every real network to be highly navigable because
navigation is not an important function of every real network. Here we
consider one example, the Pretty-Good-Privacy (PGP) web of trust
network, specifically the December 2006 snapshot and its hyperbolic
coordinates from~\cite{PaBoKr11}. These data are then processed
exactly as for all the other networks in the main text. However, as
expected, the navigation success ratio and precision metrics reported
for this network in Table~\ref{tab:pgp} are
substantially lower than for the navigable networks in the main text.

\begin{table}[H]
    \centering
      \begin{tabular}{|c|c|}
        \hline
%       \rowcolor{cyan}
        & PGP \\
        \hline
%       \rowcolor{cyan}
        Nodes & $4899$ \\
        \hline
%       \rowcolor{cyan}
        Real edges ($|R|$) & $67650$\\
        \hline
%       \rowcolor{cyan}
        NNG edges ($|M|$) & $29311$\\
        \hline
%       \rowcolor{cyan}
        True positives ($|T|$) & $6945$\\
        \hline
%       \rowcolor{cyan}
        False positives ($|F|$) & $22366$ \\
        \hline
%       \rowcolor{cyan}
        \bf  Precision ($|T|/|M|$) & $24\%$\\
        \hline
%       \rowcolor{cyan}
%        Frame edges ($|M_F|$) & $2464$\\
%        \hline
%       \rowcolor{cyan}
%        Frame true positives ($|T_F|$) & $1774$\\
%        \hline
%       \rowcolor{cyan}
%        \bf  Frame prec. ($|T_F|/|M_F|$) & $72\%$\\
%        \hline
%       \rowcolor{cyan}
        Navigation success ratio & $36\%$\\
        \hline
      \end{tabular}
    \caption{The table quantifies the relevant edge statistics showing
        the total number of edges in the core of the PGP network~$|R|$, and in its
        NNG equilibrium network~$|M|$, the number of true positive edges~$|T|=|M\cap
        R|$, the number of false positive green edges $|F|=|M\setminus R|$, and the true positive rate, or
        precision, defined as~$|T|/|M|$.
    }
    \label{tab:pgp}
\end{table}

\bibliographystyle{./paper/naturemag}

\end{document}